\documentclass[aps, prab, reprint, superscriptaddress, nofootinbib, english]{revtex4-2}

\usepackage{graphicx}
\usepackage{dcolumn}
\usepackage{amsmath}
\usepackage{amssymb}
\usepackage{mathtools}
\usepackage{xfrac}
\usepackage{hyperref}
\usepackage[binary-units]{siunitx}
\usepackage[acronym]{glossaries}

\providecommand{\wrb}{\bar{\omega}_R}
\providecommand{\rf}{\mathrm{rf}}
\providecommand{\reff}{\mathrm{ref}}
\providecommand{\wrf}{\omega_\rf}
\providecommand{\frf}{f_\rf}
\providecommand{\Real}{\mathrm{Re}}
\providecommand{\Imag}{\mathrm{Im}}
\providecommand{\der}{\mathrm{d}}
\providecommand{\mc}{\mathrm{MC}}
\providecommand{\hc}{\mathrm{HHC}}
\providecommand{\mymax}{\mathrm{max}}
\providecommand{\mymin}{\mathrm{min}}
\providecommand{\wc}{\omega_\mathrm{c}}
\DeclareMathOperator*{\argmin}{arg\,min}

\newacronym{IDFT}{IDFT}{inverse discrete Fourier transform}
\newacronym{FFT}{FFT}{fast Fourier transform}
\newacronym{llrf}{llrf}{low-level rf}
\newacronym{PI}{PI}{proportional-integral}
\newacronym{NEG}{NEG}{nonevaporable getter}
\newacronym{SD}{SD}{space-domain}
\newacronym{FD}{FD}{frequency-domain}
\newacronym{LNLS}{LNLS}{Brazilian Synchrotron Light Laboratory}
\newacronym{BPMs}{BPMs}{beam position monitors}
\newacronym{3HC}{3HC}{third-harmonic cavity}
\newacronym[plural=HHCs,firstplural=higher harmonic cavities (HHCs), longplural={higher harmonic cavities}]{HHC}{HHC}{higher harmonic cavity}
\newacronym[plural=BBRs,firstplural=broadband resonators (BBRs)]{BBR}{BBR}{broadband resonator}
\newacronym[plural=MCs,firstplural=main cavities (MCs), longplural={main cavities}]{MC}{MC}{main cavity}
\newacronym[plural=DFTs,firstplural=discrete Fourier transforms (DFTs)]{DFT}{DFT}{discrete Fourier transform}

\begin{document}

\title{Equilibrium of longitudinal bunch distributions in electron storage rings with arbitrary impedance sources and generic filling patterns}

\author{Murilo B. Alves}
\email{murilo.alves@lnls.br}
\affiliation{Brazilian Synchrotron Light Laboratory -- LNLS, Brazilian Center for Research in Energy and Materials -- CNPEM, 13083-970, Campinas, SP, Brazil.}
\affiliation{Gleb Wataghin Institute of Physics, University of Campinas -- UNICAMP, 13083-859, Campinas, SP, Brazil}

\author{Fernando H. de S\'{a}}
\email{fernando.sa@lnls.br}
\affiliation{Brazilian Synchrotron Light Laboratory -- LNLS, Brazilian Center for Research in Energy and Materials -- CNPEM, 13083-970, Campinas, SP, Brazil.}

\date{\today}

\begin{abstract}
    A new self-consistent semi-analytical method for calculating the stationary beam-induced voltage in the presence of arbitrary filling patterns and impedance sources in electron storage rings is presented. The theory was developed in space-domain with resonator wake-functions and in frequency-domain with arbitrary impedance functions. The SIRIUS storage ring parameters were used to benchmark the results, demonstrating good agreement between the two approaches and with macroparticle tracking simulations. Additionally, a different approach to simulate the beam-loading compensation of active rf cavities was investigated in frequency-domain, proving to be a more generic description than the methods generally used. The impact of broadband impedance on the longitudinal equilibrium was straightforwardly evaluated with the frequency-domain framework, without intermediate steps such as fitting broadband resonators or convolving short-range wakes with bunch distributions. Finally, a simple study of Touschek lifetime improvement with a passive higher harmonic cavity is presented.
\end{abstract}

\maketitle

\section{Introduction}
Many developments on semi-analytical methods have been made to calculate the equilibrium longitudinal bunch distributions in electron storage rings as a faster alternative to tracking codes. One of the early motivations was related to the equilibrium of a uniform filling in a double-rf system, where the effect of a passive \gls*{HHC} was investigated. In this case the beam-induced voltage has an analytical formula that can be added to the main voltage and the bunch distribution can be obtained. However, the co-dependence between these two quantities require the calculations to be iterated until convergence. The bunch profile can be accounted on the calculation of the beam-induced voltage in passive \glspl*{HHC} with a real~\cite{Byrd2001} or complex~\cite{Tavares2014} form-factor.

Synchrotron light sources often operate with nonuniform filling patterns for different reasons, for example, to allow for time-resolved experiments and to mitigate ion and coupled-bunch instabilities. Time-consuming macroparticle tracking simulations were the first attempt to study the inhomogeneous beam-loading (also referred as transient beam-loading) in the presence of passive \glspl*{HHC}. Simplified approaches considered each bunch as a pointlike macroparticle~\cite{Byrd2002, Milas2010, Ruprecht2015} or as macroparticles with Gaussian real form-factors~\cite{Phimsen2018}. Initially, semi-analytical methods were also non-self-consistent~\cite{Yamamoto2018}, assuming pointlike bunches to iteratively calculate the induced voltage. A self-consistent calculation of the inhomogeneous beam-loading was proposed in Ref.~\cite{Olsson2018}, with an iterative matrix formulation based on the linearization of the energy balance equation in the presence of a passive \gls*{HHC} modeled as a resonator. In this solution, complex form-factors were assigned for each bunch.

The problem for arbitrary filling patterns was revisited in Ref.~\cite{Warnock2020}, with the development of explicit formulas for the induced voltage of a narrowband resonator, resulting in a system of coupled Ha{\"{i}ssinski equations. Newton's method was applied to iteratively solve the problem. The theory was extended in Ref.~\cite{Warnock2021a} to include multiple resonators and an algorithm for compensating the main rf cavity beam-loading was proposed. A discussion on the effect of cavities higher order modes and short-range wakefields was presented as well. Some difficulties were reported in the convergence of the Newton iteration scheme for higher currents, when beam-induced voltages are higher. In Ref.~\cite{He2021}, phasor notation was applied to describe the induced voltage by resonators and Newton's iteration method was also employed to solve the system of equations. In this case, convergence was improved when, on each iteration,  the distributions were updated based on a linear combination of previous and present distributions, with a random coefficient as weight. The Jacobian-based iterative solution method proposed in~\cite{Warnock2020, Warnock2021a} was reappraised in Ref.~\cite{Warnock2021b}, where the equations were formulated as a fixed-point problem. The Anderson's acceleration method was introduced to enhance convergence, proving to be a robust and fast algorithm for calculating the equilibrium bunch distributions for general settings of filling patterns and resonators wakefields.

In this paper we present two semi-analytical approaches in \gls*{SD} and \gls*{FD} to obtain the beam-induced voltage. The \gls*{SD} formulation is similar in some aspects to the theory presented in Ref.~\cite{Warnock2020}. The main difference is the fact that we considered the most generic wake-function for a resonator, instead of assuming the approximated formula for large $Q$ factor. Additionally, we employed complex variables to develop the equations in \gls*{SD}, resulting in compact expressions accessible for numerical implementation and with simple interpretation. The main novelty of this work lies in the calculation of the beam-induced voltage with a \gls*{FD} framework, which allows for more general impedance models, not restricted to the resonator case. With this framework, broadband impedance and higher-order modes of rf cavities can be easily incorporated. The natural inclusion of broadband impedance sources is a very important feature, since it allows the usage of impedance models obtained from analytical and semi-analytical calculations, for which the wake-function is not available or difficult to be obtained. Moreover, the usage of impedance functions helps to establish a more realistic description of active rf cavities with a \gls*{llrf} feedback control and then evaluate its effects on the beam equilibrium. We will address the question raised in Ref.~\cite{Warnock2021a} on whether the proposed algorithm was an accurate model of the feedback mechanism, and discuss its equivalence to a particular controller type.

The paper is organized as follows: in Sec.~\ref{sec:theory} we present the theory to calculate the beam-induced voltage with two approaches. Methods to model active rf cavities and schemes of beam-loading compensation are discussed in Sec.~\ref{sec:active_rf}. In Sec.~\ref{sec:distribution_calc}, we briefly review the Ha{\"{i}ssinski equation to solve for the longitudinal bunch distributions given the beam-induced voltage. Section~\ref{sec:application} presents the application of the developed methods considering the SIRIUS storage ring parameters. Macroparticle tracking was used to benchmark the results for a nonuniform filling pattern. In the Appendix, the theory was applied to the case of uniform filling and narrowband resonator to reproduce a well-known formula for the beam-induced voltage.

\section{Beam-induced voltage\label{sec:theory}}
Throughout this report we will work with a set of global reference systems for the longitudinal coordinate~$z$ of relativistic electrons in a storage ring, with origin at the center\footnote{The center of a rf bucket is the synchronous phase considering only the energy gain by the main rf cavities and the energy loss by synchrotron radiation.} of the corresponding rf bucket~$n$ on an arbitrary turn~$r$, and $z>0$ for trailing particles. Besides, the bucket index is defined such that, if $\ell>n$, then bucket $\ell$ trails bucket $n$. In a particular coordinate system where $n=0$ and $r=0$, we can express the beam distribution, which extends through the entire real line and is one-turn periodic, as:
\begin{equation}
    \lambda_\mathrm{t}(z) = \sum_{k=-\infty}^{\infty} \lambda(z + kC_0),
    \label{eq:distribution_total}
\end{equation}
where $C_0$ is the ring circumference and $\lambda(z)$ is the one-turn distribution, given by
\begin{equation}
    \lambda(z) = \frac{1}{I_\mathrm{t}}
    \sum_{\ell=0}^{h-1} I_\ell \lambda_\ell(z - \ell \lambda_\rf), \,\,\
    \mathrm{with}\quad I_\mathrm{t} = \sum_{\ell=0}^{h-1}I_\ell > 0,
    \label{eq:distribution_oneturn}
\end{equation}
where $h$ is the harmonic number of the ring, $\lambda_\rf=C_0/h$ is the rf wavelength, $I_\ell \ge 0$ is the current of the $\ell$th bunch and $\lambda_\ell(z)$ is its distribution, which is assumed to be non-zero\footnote{The distributions are not exactly zero outside a rf period, but for electron beams it typically fall-off exponentially for $z$ sufficiently larger than the bunch length, justifying the assumption.} only for $z \in \mathcal{D} \subset [-\lambda_\rf/2, \lambda_\rf/2]$ and normalized to unity, which implies the one-turn distribution is also normalized to unity.

With this setup, the longitudinal voltage $V(z)$ induced by this current distribution under the influence of the longitudinal wake-function $W_0'(z)$ is~\cite{Ng2005}
\begin{equation}
    V(z) = -I_\mathrm{t}T_0 \int_{-\infty}^{\infty} \der z' \lambda_\mathrm{t}(z') W_0'(z-z'),
    \label{eq:wake_potential_total}
\end{equation}
where~$T_0 = C_0/c$ is the revolution period and $c$ is the speed of light. Substituting Eq.~\eqref{eq:distribution_total} into Eq.~\eqref{eq:wake_potential_total} and assuming the integral converges, we can change the order of the summation with the integral. Besides, since the choice of the turn used as origin of the coordinate system is arbitrary, we can make the following change of the integration variable $z'\to z' - kC_0$, which yields:
\begin{equation}
    V(z) = -I_\mathrm{t}T_0 \sum_{k=-\infty}^{\infty} \int_{-\infty}^{\infty} \der z' \lambda(z') W_0'(z - z' + kC_0).
    \label{eq:wake_potential_oneturn}
\end{equation}
Inserting Eq.~\eqref{eq:distribution_oneturn} in the equation above, changing the order of the summation with the integral and making the additional change of variables $z' \to z' + \ell \lambda_\rf$, we get
\begin{eqnarray}
    V(z) &=& -T_0 \sum_{k=-\infty}^{\infty} \sum_{\ell=0}^{h-1} \int_{-\infty}^{\infty} \der z' I_\ell\lambda_\ell(z') \nonumber \\
    & & \times~W_0'(z - z' + kC_0 - \ell \lambda_\rf).  \label{eq:wake_potential_onebunch}
\end{eqnarray}

Since the choice of the reference bucket was arbitrary, we could get an equivalent result using the center of the $n$th rf bucket as reference for the coordinate system. This is performed with the change of variables $z~\to~z + n\lambda_\rf$ in Eq.~\eqref{eq:wake_potential_onebunch}, which reads
\begin{eqnarray}
    V_n(z) &=& -T_0 \sum_{k=-\infty}^{\infty} \sum_{\ell=0}^{h-1} \int_{-\infty}^{\infty} \der z' I_\ell\lambda_\ell(z') \nonumber \\
           & & \times~W_0'(z - z' + kC_0 - (\ell-n)\lambda_\rf),
    \label{eq:wake_potential_nth_bunch}
\end{eqnarray}
where we introduced the notation $V_n(z) = V(z + n\lambda_\rf)$ to denote that the beam-induced voltage is calculated with the $n$th bunch as reference for the coordinate system.

\subsection{Space-domain}

The most generic longitudinal wake-function for a resonator is given by~\cite{Ng2005}:
\begin{eqnarray}
    W_0'(z) &=& 2\alpha R_s e^{-\alpha z/c}H(z) \nonumber \\
            & & \times \left[\cos(\wrb z/c) -  \frac{\alpha}{\wrb}\sin(\wrb  z/c)\right], \label{eq:wake_function_resonator}
\end{eqnarray}
where $H(z)$ is the Heaviside step function~\cite{Wolfram_Heaviside}, $\alpha > 0$ and $\wrb \ge 0$. When applied to a cavity, these two parameters are related to the quality factor $Q$ and resonant frequency $\omega_R$ by the expressions~$\alpha = \omega_R \slash 2Q$ and~$\wrb = \sqrt{\omega_R^2 - \alpha^2}$, where we note that $Q$ must be larger than $\sfrac{1}{2}$.

Let $G(z)$ be a complex function defined as
\begin{equation}
G(z) = H(z)e^{-\kappa z}~,~\kappa = (\alpha-i\wrb)/c,
\end{equation}
then the wake-function of Eq.~\eqref{eq:wake_function_resonator} can be rewritten as
\begin{equation}
    W_0'(z) = 2\alpha R_s \left\{\Real\left[G(z)\right] - \frac{\alpha}{\wrb} \Imag\left[G(z)\right]\right\}.
\end{equation}

Substituting this resonator model into Eq.~\eqref{eq:wake_potential_nth_bunch} we have
\begin{eqnarray}
    V_n(z) &=& -2\alpha R_s T_0 \sum_{k=-\infty}^{\infty}\sum_{\ell=0}^{h-1} \int_{-\infty}^{\infty}{\der z' I_\ell \lambda_{\ell}(z')} \nonumber \\
           & & \times~\left\{\Real\left[G(\zeta_{k\ell n})\right] - \frac{\alpha}{\wrb} \Imag\left[G(\zeta_{k \ell n})\right]\right\},
\end{eqnarray}
where~$\zeta_{k \ell n} = z - z' + kC_0 - (\ell-n)\lambda_\rf$ was introduced.

Since the distributions $\lambda_\ell(z)$ are real, if we define the following complex function
\begin{equation}
    K_n(z) = \sum_{k=-\infty}^{\infty}\sum_{\ell=0}^{h-1} \int_{-\infty}^{\infty}{\der z' I_\ell \lambda_{\ell}(z')} G(\zeta_{k\ell n}),
    \label{eq:kn_definition}
\end{equation}
then the beam-induced voltage for the $n$th bunch can be compactly written as
\begin{equation}
    V_n(z) = -2\alpha R_s T_0\left\{\Real\left[K_{n}(z)\right] - \frac{\alpha}{\wrb} \Imag\left[K_{n}(z)\right]\right\}.
    \label{eq:wake_voltage_compact}
\end{equation}

The causality property of the wake-function is encoded in the function~$G(z)$ by means of the Heaviside step function. Nevertheless, this property should be explicitly manifested in the integration and summation limits to further simplify our expressions. This can be done with the following arguments.
\begin{itemize}
\item The bunch distributions $\lambda_\ell(z)$ are assumed to be zero outside the interval $[-\lambda_\rf/2, \lambda_\rf/2]$, so the limits of integration in Eq.~\eqref{eq:kn_definition} could be restricted to this range. With this consideration, we note that $\zeta_{k\ell n} < 0$ for $k<0$ and the summation over turns can be reduced to non-negative values of $k$.
\item For $\ell=n$ and $k=0$, i.e., the self-induced voltage of a particular bunch in the present turn, causality is obeyed when $z'\leq z$, limiting the integration domain to $(-\infty, z)$.
\item Taking $\ell < n$, which means the source bunch $\ell$ leads the trailing bunch $n$, then $k=0$ should be considered in the summation. For $\ell > n$ only $k > 0$ should be accounted.
\end{itemize}

Applying these considerations into Eq.~\eqref{eq:kn_definition} results to
\begin{equation}
    K_n(z) = e^{-\kappa z} \left(I_n S_n(z)+ \sum_{\ell=0}^{h-1}M_{n\ell}A_{n\ell} I_\ell\right)
    \label{eq:k_function}
\end{equation}
where $S_n(z)$ is related to the effect of the bunch $n$ on itself in the present turn ($k=0$), given by
\begin{equation}
    S_n(z) \coloneqq \int_{-\infty}^{z}{\der z' \lambda_n(z')}e^{\kappa z'},
\end{equation}
the terms $A_{n\ell}$ are given by
\begin{equation}
    A_{n\ell} \coloneqq \sum_{k=\left\{\substack{0,\, \ell < n \\1,\, \ell \geq n}\right.}^{\infty} \nu^k =
    \begin{cases}\frac{1}{1-\nu}, & \ell < n  \\ \frac{\nu}{1-\nu}, & \ell \geq n\end{cases},
\end{equation}
where $\nu = e^{-\kappa C_0}$, and $M_{n\ell}$ is defined as
\begin{equation}
    M_{n\ell} \coloneqq e^{\kappa(\ell - n)\lambda_\rf}\int_{-\infty}^{+\infty}{\der z'\lambda_{\ell}(z')}e^{\kappa z'},
\end{equation}
which depends on the bilateral Laplace Transform~\cite{Wolfram_Bilateral_Laplace} of the bunch distribution evaluated at $-\kappa$, which in turn can be identified as~$S_\ell(z)$ in the limit that~$z$ tends to infinity. Moreover, since $\lambda_\ell(z)$ is zero outside the interval $\mathcal{D}\subset[-\lambda_\rf/2, \lambda_\rf/2]$, then the following is valid:
\[
    \lim_{z \to \infty} S_\ell(z) = S_\ell(\lambda_\rf/2).
\]

Equation~\eqref{eq:k_function} has a straightforward numerical implementation when we consider a uniformly discretized $z$ domain for each bucket:
\begin{equation}
z_j = \frac{\lambda_\rf}{a}\left(\frac{j}{N} - \frac{1}{2}\right),\quad j=0, \dots, N-1,
\label{eq:z_discretization}
\end{equation}
where $1 \leq a \in \mathbb{R}$ and $N \in \mathbb{N}$ should be chosen appropriately, depending on the typical bunch length, resonator frequency and damping rate.

Numerical problems related to floating-point overflow may arise in calculations when high-frequency ($\omega_R \gtrsim \SI{100}{\giga\hertz}$) low-$Q$ resonators are involved, due to the exponential with positive real argument in $M_{n\ell}$ when $\ell>n$. In these cases it is recommended to use the identity $g(x)e^x = e^{\log(g(x))+ x}$ to avoid such problems, thus $M_{n\ell}A_{n\ell} = e^{\log(M_{n\ell}) + \log(A_{n\ell})}$ and the term $-\kappa C_0$ compensates $\kappa(\ell-n)\lambda_\rf$. For even higher frequencies, the calculation of $S_n(z_j)$ may have similar issues and the same approach can be used for the integrand. Besides, the integral can be suitably truncated once the integrand approaches zero, which will generally be the case, given that the bunch distributions fall-off faster than the exponential term~$e^{\kappa z'}$.

It is also straightforward to include an arbitrary number of resonators in the calculations, since, by linearity, the induced voltages for each resonator can be added. However, the calculation time grows linearly with the number of resonators, given that all quantities from Eq.~\eqref{eq:wake_voltage_compact} and Eq.~\eqref{eq:k_function} must be re-evaluated as the resonator parameters change.

It is possible to further simplify\footnote{The calculation steps are: (i) insert the explicit expressions for $M_{n\ell}$ and $A_{n\ell}$ into Eq.~\eqref{eq:k_function}; (ii) rewrite $\nu = e^{-\kappa C_0} = e^{-\kappa h \lambda_\rf}$; (iii) separate the terms $\ell<n$ and $\ell \geq n$ in the summation; (iv) for the sum with $\ell\geq n$, re-index the summation variable $\ell'=\ell-h$; (v) use the property $I_{\ell\pm h} = I_{\ell}$ and $\lambda_{\ell\pm h}(z) = \lambda_{\ell}(z)$; (vi) identify that both summands are equal and unify the sums; (vii) re-index the summation variable $\ell'=n-\ell$.} Eq.~\eqref{eq:k_function} for a convenient interpretation of the beam-induced voltage:
\begin{equation}
    K_n(z) = e^{-\kappa z} \left(I_n S_n(z) + \sum_{\ell=1}^{h}\frac{\nu^{\ell/h}}{1-\nu}S_{n-\ell}(\lambda_\rf/2)I_{n-\ell}\right).
    \label{eq:k_function2}
\end{equation}
In this expression we note that the voltage acting on the $n$th bunch is the sum of its own action on the current turn and the effect of previous passages of all bunches, including itself, with an appropriate phase and decay factor. It also facilitates to check the continuity of the voltage between adjacent buckets: $K_n(\lambda_\rf/2)~=~K_{n+1}(-\lambda_\rf/2)$. Interestingly, Eq.~\eqref{eq:k_function2} is free from the numerical issues related to positive arguments in exponential.

\subsection{\label{subsec:impedance_calc}Frequency-domain}
An arbitrary longitudinal wake-function $W_0'(z)$ is related to a longitudinal impedance $Z(\omega)$ by the inverse Fourier transform~\cite{Ng2005}:
\begin{equation}
    W_0'(z) = \dfrac{1}{2\pi}\int_{-\infty}^{\infty} \der\omega Z(\omega)e^{-i\omega z/c}.
\end{equation}

Inserting this relation into Eq.~\eqref{eq:wake_potential_oneturn} reads
\begin{eqnarray}
 V(z) &=& -\dfrac{I_\mathrm{t}T_0}{2\pi}\sum_{k=-\infty}^{\infty} \int_{-\infty}^{\infty}{\der z'}\int_{-\infty}^{\infty} {\der\omega}~\lambda(z') \nonumber \\
      & & \times~Z(\omega)e^{-i\omega(z-z'+ kC_0)/c}.
\end{eqnarray}

Rearranging the exponential terms, we can apply the Poisson sum formula to the summation over turns:
\begin{equation}
\sum_{k=-\infty}^{+\infty}e^{-ik\omega T_0} = \omega_0 \sum_{p=-\infty}^{+\infty}\delta(\omega+ p\omega_0),
\end{equation}
where~$\delta(\cdot)$ is the Dirac delta distribution and~$\omega_0=2\pi/T_0$. With this change, the integral over $\omega$ can be easily performed. The expression simplifies to
\begin{eqnarray}
V(z) &=& -I_\mathrm{t}\sum_{p=-\infty}^{+\infty}Z^\ast(p\omega_0)e^{ip\omega_0z/c} \nonumber \\
     & &\times~\int_{-\infty}^{\infty}\der z'\lambda(z')e^{-ip\omega_0z'/c},
\label{eq:voltage_impedance_partial}
\end{eqnarray}
where we used the property $Z(-\omega) = Z^\ast (\omega)$, with~${}^\ast$ denoting the complex conjugate.

The numerical implementation of Eq.~\eqref{eq:voltage_impedance_partial} requires truncation of the infinity sum over harmonics~$p$. Two alternative approaches will be presented to properly select the harmonics. One is based on the \gls*{DFT} of the one-turn distribution, which considers all harmonics up to a specific threshold. The other is based on a selection of the most relevant harmonics, depending on the filling pattern and impedances under consideration. While the first method is generally much faster, since it benefits from the use of the \gls*{FFT} algorithm, the second one is better suited when the impedance is composed of a few narrowband peaks.

\subsubsection{Implementation with DFT}\label{subsub:dft_implementation}
Consider the case of the discretized~$z$-coordinate from Eq.~\eqref{eq:z_discretization} for each bucket with $a=1$, thus $z$ covers one rf period with~$N$ points. This $z$-coordinate can be used for all $h$ buckets in one-turn, concatenating it $h$ times to form a discretized coordinate with $hN$ elements, extending to the domain~$\mathcal{T} = \left[-\lambda_\rf/2,C_0-\lambda_\rf/2\right]$ in which the one-turn distribution~$\lambda(z)$ is defined. For a sufficiently small spacing~$\Delta z=\lambda_\rf/N$, we can approximate the integral from Eq.~\eqref{eq:voltage_impedance_partial} by quadrature to:
\begin{eqnarray}
\int_{\mathcal{T}}\der z'\lambda(z')e^{-ip\omega_0z'/c} &\approx& e^{\pi i p/h}\Delta z \nonumber \\
                                                        &       & \times~\sum_{j=0}^{hN-1}\lambda(z_j)e^{-2\pi i \frac{pj}{hN}}.
\label{eq:quadrature}
\end{eqnarray}

To establish notation, the \gls*{DFT} of a sequence of~$N$ real numbers~$\mathbf{x}~=~\left[x_0, x_1, \ldots, x_{N-1}\right]$ and the \gls*{IDFT} are defined as
\begin{eqnarray}
    X_k = \mathcal{F}\left\{\mathbf{x}\right\}_k &=& \sum_{n=0}^{N-1} x_n e^{-2\pi i k n/N},  \, \, \, \forall~k \in \mathbb{Z} \\
    x_n = \mathcal{F}^{-1}\left\{\mathbf{X}\right\}_n &=& \frac{1}{N}\sum_{k=-\lfloor (N-1)/2\rfloor}^{\lfloor N/2\rfloor} X_k e^{2\pi i kn/N}
\end{eqnarray}
where $\lfloor\cdot\rfloor$ is the floor operation. Note that, even though the \gls*{DFT} is defined for all $k \in\mathbb{Z}$, only a sequence of $N$ consecutive terms are needed to compute the \gls*{IDFT}.

With those definitions, the summation in Eq.~\eqref{eq:quadrature} can be identified as the \gls*{DFT} of the sequence~$\left[\lambda(z_j)\right]$. Hence
\begin{equation*}
\int_{\mathcal{T}}\der z'\lambda(z')e^{-ip\omega_0z'/c} \approx  e^{\pi i p/h}\Delta z\mathcal{F}\left\{\lambda(z)\right\}_p.
\end{equation*}

Applying this result to Eq.~\eqref{eq:voltage_impedance_partial} for the discretized coordinate $z_n$ we obtain
\begin{equation}
    V(z_n) = -I_\mathrm{t}\Delta z\sum_{p=-\infty}^{\infty}Z^\ast(p\omega_0)\mathcal{F}\left\{\lambda(z_j)\right\}_p e^{2\pi i \frac{pn}{hN}},
    \label{eq:wake_voltage_impedance_dft_partial}
\end{equation}
where the phase terms $e^{\pm\pi i p/h}$ nicely canceled each other.

Considering that the grid spacing was properly chosen, the minimum and maximum frequencies calculated by the \gls*{DFT}, $-\lfloor (hN-1)/2\rfloor\omega_0$ and $\lfloor hN/2\rfloor\omega_0$, should be large enough so the bunch distribution does not have any significant contribution from frequencies outside this interval. With that in mind, we can truncate the infinite sum over $p$ in Eq.~\eqref{eq:wake_voltage_impedance_dft_partial} to the limits of the \gls*{IDFT}, yielding
\begin{equation}
    V(z_n) = -I_\mathrm{t}C_0\mathcal{F}^{-1}\left\{Z^\ast(p\omega_0)\mathcal{F}\left\{\lambda(z_j)\right\}_p\right\}_n
    \label{eq:wake_voltage_impedance_dft}
\end{equation}
where $hN\Delta z = C_0$ was applied since $\Delta z = \lambda_\rf/N$.

\subsubsection{Relevant Harmonics Selection}
Starting from Eq.~\eqref{eq:voltage_impedance_partial} we can apply the definition of the one-turn distribution from Eq.~\eqref{eq:distribution_oneturn} and rearrange the exponential terms to get
\begin{eqnarray}
    V_n(z) &=& -\sum_{p=-\infty}^{+\infty}Z^\ast(p\omega_0)e^{ip\omega_0(z + n\lambda_\rf)/c} \nonumber \\
           & & \times~\sum_{\ell=0}^{h-1}\int_{-\infty}^{\infty}\der z'I_\ell\lambda_\ell(z')e^{-ip\omega_0(z'+\ell\lambda_\rf)/c}.
    \label{eq:wake_voltage_impedace_partial_2}
\end{eqnarray}

Observing that the terms in the sum over $p$ become their conjugate for $p \to -p$ and considering that only a subset $\mathcal{P}\subset[0, \infty)$ will be kept in the sum, Eq.~\eqref{eq:wake_voltage_impedace_partial_2} can be transformed into:
\begin{eqnarray}
    V_n(z) &=& -2\Real\left[\sum_{p\in\mathcal{P}}Z^\ast( p\omega_0)e^{ip\omega_0z/c}e^{2\pi i pn/h}\right.    \nonumber \\
           & & \times~\left.\sum_{\ell=0}^{h-1} I_{\ell}\hat{\lambda}_{\ell}^\ast(p\omega_0)e^{-2\pi i p\ell/h}\right]
    \label{eq:wake_voltage_impedance_2}
\end{eqnarray}
where
\begin{equation}
    \hat{\lambda}_\ell(\omega) \coloneqq \int_{-\infty}^{+\infty}\der z' \lambda_\ell(z) e^{i\omega z'/c}
    \label{eq:fourier_transform}
\end{equation}
is the Fourier transform of the longitudinal distribution.

 The determination of a subset $\mathcal{P}$ that keeps the truncation error small can be done as follows: (i) calculate the \gls*{DFT} of the filling pattern $\mathbf{I}_b~=~\left[I_0, I_1, \ldots, I_{h-1}\right]$, (ii) sample the impedance at harmonics $p \in \mathcal{P}_{\max} = [0, 1, \ldots, p_{\max}]$, where $p_{\max}$ must be larger than the maximum relevant frequency, depending on the distribution and impedance under consideration, (iii) determine the subset
\begin{equation}
    \label{eq:modes_selection_criteria}
    \mathcal{P} = \left\{p \in \mathcal{P}_{\max} \mid \xi(p) \geq \xi_{\min}\right\},
\end{equation}
where $\xi(p) = \lvert Z\left(p\omega_0\right) \mathcal{F}\left\{\mathbf{I}_b\right\}_p \rvert$ and $\xi_{\min}\in\mathbb{R}$ is a minimum threshold. Including the filling pattern frequency spectrum is important because for arbitrary fills the beam samples the impedance at revolution harmonics and the most relevant modes might be non-trivial. The threshold can be set as~$\xi_{\min}=\max\left[\xi(\mathcal{P}_\mymax)\right]\varepsilon$, where~$\varepsilon$ can be made as small as needed so that no considerable change is observed in the equilibrium solution.

One drawback of this implementation, compared to the \gls*{DFT} approach, is that the Fourier transform of $h$ bunch distributions for all $p \in \mathcal{P}$ must be evaluated via numerical integration. This process has a time complexity of $\mathcal{O}(hN\lvert \mathcal{P} \rvert)$, where $\lvert \mathcal{P} \rvert$ denotes the cardinality of $\mathcal{P}$, while the computation of the one-turn distribution \gls*{DFT} has a complexity of $\mathcal{O}(hN\log(hN))$. On the other hand, the harmonics selection approach allows a free choice of the discretization interval (any $a\geq 1$ in Eq.~\eqref{eq:z_discretization}), which can improve accuracy for some cases. Note that the calculations with the \gls*{SD} framework has a time complexity of $\mathcal{O}(hN N_\mathrm{R})$, where $N_\mathrm{R}$ is the number of resonators\footnote{Time complexity discussions are appropriate for the limit of large numbers. In this case, for large values of $N$, $h$, $N_\mathrm{R}$ and $\lvert \mathcal{P} \rvert$.}.

\section{Active \lowercase{rf} cavities\label{sec:active_rf}}
For active rf cavities the total voltage inside the cavity, $V_\mathrm{t}$, is the sum of the generator voltage, $V_\mathrm{g}$, supplied by an external power source, and the beam-induced voltage, $V_\mathrm{b}$, commonly called beam-loading. In general, $V_\mathrm{g}$ is varied through the action of feedback loops such that $V_\mathrm{t}$ is kept close to a constant reference value, $V_\mathrm{r}$, in a narrow bandwidth around the center frequency $\wc$. We will assume in the next steps that $\wc$ is a multiple of the revolution frequency, but not necessarily a multiple of the main rf frequency $\wrf$.

In tracking simulations it is common to simulate the beam-loading compensation scheme with realistic models of the feedback system~\cite{Berenc2015, Rivetta2007}. However, in equilibrium simulations the time-dependence of the system is neglected and very idealized models are generally used, which do not take into account the system delays or the effect of the system on neighboring revolution harmonics. In this section we will discuss some conventional methods to simulate the beam-loading compensation and present an approach, with straightforward implementation in the \gls*{FD} framework, that allows for realistic simulation of \gls*{llrf} feedback systems that are typically used to control the voltage in
active cavities.

\subsection{Least squares minimization}
A scheme to calculate the generator voltage parameters is to minimize
the following difference for each bucket
\begin{equation}
    \chi_n^2 = \int_{-\lambda_\rf/2}^{\lambda_\rf/2}{\der z{\left[V_{\mathrm{g}, n}(z) + V_{\mathrm{b}, n}(z) - V_{\text{r}, n}(z)\right]}^2},
    \label{eq:chi2_generator}
\end{equation}
where $V_{b, n}(z)$ is the beam-loading voltage for bucket $n$, which can be calculated using the impedance or wake-function model for the cavity and the techniques presented in Sec.~\ref{sec:theory}. The reference and generator voltages are given by
\begin{eqnarray*}
    V_{\mathrm{r}, n}(z) &=& \Real\left[\hat{V}_\mathrm{r}\,e^{i\wc (z+n\lambda_\rf)/c}\right], \\
    V_{\mathrm{g}, n}(z) &=& \Real\left[\hat{V}_\mathrm{g}\,e^{i\wc (z+n\lambda_\rf)/c}\right],
\end{eqnarray*}
where we made use of the notation introduced in Eq.~\eqref{eq:wake_potential_nth_bunch} to take bucket $n$ as reference. Note that, if $\wc$ is a multiple of $\wrf$, then both voltages have the same phase relation for all buckets.  The phasors are defined by the respective amplitudes and phases with:
\begin{equation}
\hat{V}_\mathrm{r} = V_\mathrm{r} e^{i(\pi/2-\phi_\mathrm{r})}
\,\,\mathrm{and}\,\,
\hat{V}_\mathrm{g} = V_\mathrm{g} e^{i(\pi/2-\phi_\mathrm{g})},
\end{equation}
where $\phi_\mathrm{r}$ is the reference phase.

The minimization of Eq.~\eqref{eq:chi2_generator} with respect to the amplitude and phase of the generator voltage can be rewritten as a linear problem with
\begin{equation*}
V_{\mathrm{g}, n}(z)=A\sin\left(\frac{\wc}{c}(z+n\lambda_\rf)\right)+B\cos\left(\frac{\wc}{c}(z+n\lambda_\rf)\right)
\end{equation*}
where $A=V_\mathrm{g}\cos(\phi_\mathrm{g})$ and $B~=~V_\mathrm{g}\sin(\phi_\mathrm{g})$ are free parameters. With this setup, the minimization problem can be formally written as
\begin{equation}
    (\tilde{A}, \tilde{B}) = \argmin\limits_{(A, B)} \sum_{\ell=0}^{h-1}\chi_\ell^2.
    \label{eq:least_squares_minimization}
\end{equation}

The numerical implementation of this method is straightforward and will not be presented here\footnote{An equivalent approach was used in Ref.~\cite{Warnock2021a}, with the derivation of analytic expressions for the Jacobian taking into account the beam response in front of the changing parameters of the generator. In our implementations we noted that a simple numeric estimation of the Jacobian, without accounting for the changes in $V_\mathrm{b}$ were enough to reach convergence.}.

\subsection{Phasor compensation}
The beam-loading voltage phasor at the center frequency $\wc$ can be calculated as
\begin{equation}
    \hat{V}_\mathrm{b}(\wc) = \frac{2}{C_0}\sum_{\ell=0}^{h-1}\int_{-\lambda_\rf/2}^{\lambda_\rf/2}{\der z V_{\mathrm{b}, \ell}(z)e^{-i\wc z/c}},
    \label{eq:beam_loading_phasor}
\end{equation}
where we observe that the combination of the sum with the integral is equivalent to an integration along the whole storage ring, which guarantees that only the harmonic $p = \wc/\omega_0$ will influence the phasor. Note that the numerical implementation of Eq.~\eqref{eq:beam_loading_phasor}, as well as Eq.~\eqref{eq:chi2_generator}, requires $a=1$ in the discretization defined by Eq.~\eqref{eq:z_discretization}. In that way, no other harmonic of the beam-loading influences the compensation scheme.

With the phasors for induced and reference voltages calculated, the generator voltage phasor can be set as:
\begin{equation}
    \hat{V}_\mathrm{g} = \hat{V}_\mathrm{r} - \hat{V}_\mathrm{b}(\wc).
\end{equation}
It is possible to demonstrate that this method is equivalent to the least squares minimization method presented previously.

\subsection{\label{subsec:closed_loop_impedance}Closed-loop impedance}
While the two previous methods of calculating the beam-loading of active cavities can be implemented either in \gls*{SD} or \gls*{FD}, the next one is particular for the \gls*{FD} approach. In this framework, it is possible to set $V_\mathrm{g}=V_\reff$ and simulate the compensation by using the effective impedance of the cavity seen by the beam in the presence of a \gls*{llrf} control loop~\cite{Rivetta2007, Baudrenghien2017, Karpov2019}:
\begin{equation}
    Z_{\text{cl}}(\omega) =\frac{V_\mathrm{t}(\omega)}{I_\mathrm{b}(\omega)} = \frac{Z(\omega)}{1 + T(\omega)Z(\omega)},
\end{equation}
where $Z(\omega)$ is the open-loop impedance of the cavity, which can be modelled as the impedance of an equivalent RLC circuit~\cite{Ng2005}
\begin{equation}
    Z(\omega) = \dfrac{R_\mathrm{s}}{1+ iQ\left(\dfrac{\omega_R}{\omega} - \dfrac{\omega}{\omega_R}\right)},
    \label{eq:rlc_impedance}
\end{equation}
where $R_s$ and $Q$ are the cavity shunt impedance and quality factor. $T(\omega)$ is the rf plant transfer function apart from the cavity impedance. Thus, the overall open-loop transfer function is $L(\omega) = T(\omega)Z(\omega)$.

A simple model for $Z_\mathrm{cl}(\omega)$ is obtained by letting $T(\omega) \to \delta(\omega-\wc)$, which assumes perfect compensation of the beam-loading component at the control loop frequency, since $Z_\mathrm{cl}(\wc)=0$, and the loop is transparent for all other frequencies ($Z_\mathrm{cl}(\omega)=Z(\omega), \forall\, \omega \neq \wc$). It can be shown that this model is equivalent to the methods presented previously if only static beam-loading is considered\footnote{Dynamic beam-loading will also contain frequencies that are not multiple of revolution harmonics, such as multiples of the synchrotron frequency, which will contribute to the calculation of the cost function defined in Eqs.~\eqref{eq:chi2_generator} and~\eqref{eq:least_squares_minimization} and the phasor of Eq.~\eqref{eq:beam_loading_phasor}, since these components are not orthogonal to the $\wc$ component in the integration and summation domains. On the other hand, they would not be accounted in the generator voltage in this closed-loop impedance method. This scenario, however, is outside the scope of this work, since we are concerned only with the equilibrium state.}.

To simulate a more realistic feedback, one can consider the following model for the rf plant transfer function:
\begin{equation}
    T(\omega) = C(\omega)\mathcal{K}e^{-i\left(\omega\tau_\mathrm{d} - \phi\right)},
    \label{eq:rf_plant_model}
\end{equation}
which consists of overall gain $\mathcal{K}$ and delay $\tau_\mathrm{d}$, a controller $C(\omega)$ and a phase $\phi$ that can be adjusted such that $\phi = \wc\tau_\mathrm{d}$, i.e., the overall phase is zero at the control frequency $\wc$.

Considering a purely proportional feedback, $C(\omega)=k_\mathrm{p}$, it is possible to show that the flat-response for the closed-loop system is obtained by setting the feedback gain to~\cite{Boussard1985}:
\begin{equation}
    \frac{1}{k_\mathrm{p, f}\mathcal{K}} = \frac{2}{\pi}\left(\frac{R}{Q}\right)\wrf\tau_{\mathrm{d}},
    \label{eq:flat_response_gain}
\end{equation}
which will result in $Z_\mathrm{cl}(\wc) = 1/(k_\mathrm{p, f}\mathcal{K})$.

The \gls*{PI} controller is widely used in digital \gls*{llrf} systems. Generally these systems down-convert the rf signal and then adjust the generator voltage amplitude and phase by applying the control law on the digitized quadrature components of the signal in baseband. There are several techniques to accomplish this, whose detailed modelling and description is beyond the scope of this work. However, a very simplified model of this type of controller, that does not take into account nonlinear effects nor the analog-to-digital and digital-to-analog conversions, is presented below:
\begin{equation}
    C(\omega) = k_\mathrm{p} + \frac{k_\mathrm{i}}{i(\omega-\wc)},\,\, \omega \geq 0
\end{equation}
where $k_\mathrm{i}$ is another free parameter. The domain restriction to non-negative frequencies and the term $(\omega - \wc)$ are related to the up-conversion of the integrator applied in baseband. The relation $C(-\omega) = C^\ast(\omega)$ must be used to evaluate the transfer function for negative frequencies. This controller model also strongly suppress the beam-loading at the control frequency, but, differently from the previous methods, it allows for the evaluation of the control system impact on neighboring revolution harmonics. In practice this effect will largely depend on the specificities of each system, such as the strength of $k_\mathrm{p}$ and $k_\mathrm{i}$, the filters that are used to limit the bandwidth of the controller or even other factors such as the unmodeled dynamics. However, in principle it should be possible to improve the model of the rf plant of interest and use the corresponding closed-loop impedance on the \gls*{FD} framework to have a reasonable characterization of the effect of \gls*{llrf} control loop on equilibrium parameters.

\section{Equilibrium bunch distributions\label{sec:distribution_calc}}
The equilibrium longitudinal distribution of bunch $n$, $\lambda_n(z)$, in an electron storage ring is given by the Ha{\"{i}ssinski equation~\cite{Haissinski1973}:
\begin{eqnarray}
    \lambda_n(z) &=& A_n \exp{\left(-\frac{\Phi_n(\lambda; z)}{\alpha_c \sigma_\delta^2}\right)}, \,\,\mathrm{with}\label{eq:haissinski} \\
    \Phi_n(\lambda; z) &=& -\frac{1}{E_0C_0}\int_{0}^z {\der z' \left[eV_{\mathrm{t},n}(\lambda; z') - U_0\right]},
\end{eqnarray}

where $\lambda=\lambda(z)$ is the equilibrium one-turn distribution, given by Eq.~\eqref{eq:distribution_oneturn}, $\sigma_\delta$ is the equilibrium relative energy spread, $A_n$ is a normalization constant, $\alpha_c$ is the momentum compaction factor, $E_0$ is the ring nominal energy, $e>0$ is the elementary charge, $U_0$ is the energy loss per turn from synchrotron radiation and
\[
    V_{\mathrm{t},n}(\lambda; z) = V_{\text{g}, n}(z) + V_{\mathrm{b}, n}(\lambda; z)
\]
is the total voltage, written in terms of the generator voltage and the beam-induced voltage at the $n$th bunch, given by Eq.~\eqref{eq:wake_potential_nth_bunch}.

Equation~\eqref{eq:haissinski} can be solved numerically by different methods, for example, calculating the self-consistent distribution with fixed-point algorithms~\cite{Warnock2021b} or with a Jacobian-based algorithm as Newton's method~\cite{Warnock2020, He2021, Warnock2021a}. In this work we employed Anderson's algorithm to enhance fixed-point iterations, and we refer to Ref.~\cite{Warnock2021b} and its references for more information on this subject.

To determine the convergence of the iterative process, it is convenient to define a functional $\Delta: \mathbb{R}^{hN} \times \mathbb{R}^{hN} \to \mathbb{R}$ ($N$ is the number of points on the~$z$ grid) that measures the difference of two one-turn distributions $\lambda(z')$ and $\eta(z')$ by:
\begin{eqnarray}
    \Delta\left(\lambda, \eta\right) &=& \max_n \Delta_n\left(\lambda_n, \eta_n\right), \,\, \mathrm{with} \\
    \Delta_n\left(\lambda_n, \eta_n\right) &=& \int_{-\lambda_{\rf}/2}^{\lambda_{\rf}/2}\der z'|\lambda_n(z')-\eta_n(z')|.
\end{eqnarray}

The iterative process can be terminated at iteration $k$ if the last two distributions are sufficiently close, i.e., if~$\Delta \left(\lambda^{(k)}, \lambda^{(k-1)}\right) < \Delta_\mymin$ is satisfied, where $\Delta_\mymin$ is a convergence parameter.

\section{Applications for SIRIUS\label{sec:application}}
We will discuss in this section some interesting cases to benchmark the formulas presented previously and also to highlight the advantages of using the \gls*{FD} over the \gls*{SD} approach. The numerical implementation was carried out in python3 and the code is open to access~\cite{pycolleff_github}. The solution of the Ha{\"{i}ssinski equation was computed with fixed-point iterations accelerated by Anderson's algorithm~\cite{Warnock2021b}. Gaussian distributions for all bunches were always taken as the initial condition. The results from semi-analytical methods for a nonuniform filling pattern were benchmarked against a macroparticle tracking code that was also implemented in python3~\cite{pycolleff_github}. The implementation is similar to the one described in Ref.~\cite{Yamamoto2018}, computing the time evolution of the longitudinal dynamic variables of several macroparticles in each rf bucket in the presence of resonator wakefields, with the effects of radiation damping and quantum excitation taken into account. We followed the strategy of tracking just a few macroparticles per bunch at the initial turns to speed up computing time. Then the number was gradually increased by oversampling the existing particles with a small random variation of their coordinates.

In all semi-analytical simulations reported in this work we used $N=2001$ and $a=1$ in~Eq.~\eqref{eq:z_discretization} to discretize the $z$ coordinate. The relaxation parameter in Anderson's acceleration method was fixed at the value of \num{0.1} and provided fast convergence for all evaluated cases. It was sufficient to consider a linear combination of three previous distributions to update the distribution for the next iteration. In the notation established in Ref.~\cite{Warnock2021b}, we set $m=3$ and $\beta_k = 0.1$. It was checked that the convergence criteria of $\Delta_\mymin~=~10^{-8}$ was a good trade-off to obtain a reliable fixed-point solution while reducing the total number of iterations. Regarding the tracking simulations, we adhered to the following schedule for increasing the number of particles: \num{100} particles per bunch in the first \num{50000} turns; then \num{1000} particles per bunch in the following \num{20000} turns; and \num{10000} particles per bunch in the final \num{10000} turns. All calculations were performed on the same personal computer with quite modest hardware configurations: an 8th generation Intel Core i7 processor, \SI{32}{\giga\byte} of RAM memory and no graphics processing unit capabilities.

The SIRIUS storage ring, a fourth-generation synchrotron light source built and operated by the \gls*{LNLS} in Campinas, Brazil~\cite{SIRIUS_IPAC23}, was used to exemplify the application of the formulas developed in the previous sections. The main parameters for the machine are described in Table~\ref{tab:sirius_params}.
\begin{table}
\caption{Main parameters for SIRIUS storage ring}
\label{tab:sirius_params}
\begin{ruledtabular}
    \begin{tabular}{lcc}
        Parameter & Symbol & Value \\
        \hline
        Energy &$E_0$ & $\SI{3}{\giga\electronvolt}$  \\
        Nominal current &$I_\mathrm{t}$ & $\SI{350}{\milli\ampere}$ \\
        Circumference &$C_0$  & $\SI{518.39}{\meter}$ \\
        Harmonic number &$h$ &  864 \\
        Momentum compaction factor &$\alpha$ & $\SI{1.645e-4}{}$  \\
        Energy loss per turn (with IDs) &$U_0$ & $\SI{870}{\kilo\electronvolt}$  \\
        Relative energy spread &$\sigma_\delta$ & $\SI{8.436e-4}{}$  \\
        Natural rms bunch length &$\sigma_{z, 0}$ & $\SI{2.6}{\milli\meter}$ \\
        rf frequency &$\frf$ &  $\SI{499.667}{\mega\hertz}$   \\
        Number of \glspl*{MC} & $N_\mc$ & 2 \\
        \glspl*{MC} total voltage &$N_\mc V_\mc$ & $\SI{3.0}{\mega\volt}$ \\
        \gls*{MC} geometric factor &$(R/Q)_\mc$ & $\SI{89}{\ohm}$ \\
        \gls*{MC} unloaded quality factor &$Q_{0, \mc}$ & $\SI{2e9}{}$ \\
        External quality factor &$Q_\mathrm{ext}$ & $\SI{1.58e5}{}$ \\
        Cavity coupling factor  &$\beta_{\mathrm{c}}$& \SI{12657}{} \\
        \gls*{MC} detuning\footnote{Calculated to minimize the
        reflected power.} &$\Delta f_\mc$& $\SI{-4.9}{\kilo\hertz}$ \\
        Number of \glspl*{HHC} & $N_\hc$ & 1 \\
        \gls*{HHC} rf harmonic & $\omega_\hc/\omega_\rf$ & 3 \\
        \gls*{HHC} geometric factor &$(R/Q)_\hc$ & $\SI{87.5}{\ohm}$ \\
        \gls*{HHC} quality factor &$Q_{0, \hc}$ & $\SI{4e8}{}$ \\
        \gls*{HHC} flat-potential voltage ratio & $V_\hc/N_\mc V_\mc$ & 0.317 \\
        \gls*{HHC} detuning\footnote{To provide the flat-potential harmonic voltage in uniform filling.} & $\Delta f_\hc$ & $\SI{45}{\kilo\hertz}$
    \end{tabular}
\end{ruledtabular}
\end{table}
We could not compare the presented simulated results with experimental data since SIRIUS storage ring is operating with a temporary normal conducting PETRA 7-cell rf cavity and the \gls*{3HC} is not installed yet. The definitive rf system for SIRIUS will have two superconducting CESR-B \glspl*{MC} and a superconducting passive \gls*{3HC}, according to the parameters presented in Table~\ref{tab:sirius_params}.

\subsection{Benchmarking\label{subsec:benchmark}}

\subsubsection{Uniform filling}
First we simulated the case of nominal current in uniform filling with a superconducting passive \gls*{3HC} modeled as a resonator, following the parameters from Table \ref{tab:sirius_params}. Under these conditions, all bunches are equivalent and there is an analytical formula for the beam-induced voltage in the resonator, given by Eq.~\eqref{eq:analytic_voltage}, which was used to benchmark the calculations. We considered that formula in the fixed-point iteration to solve for the corresponding longitudinal distribution, which we will denote as~$\lambda_{\text{A}}(z)$. The resulting distribution centroid is $\SI{-0.23}{\milli\meter}$ and the rms bunch length is $\SI{11.88}{\milli\meter}$, yielding a bunch lengthening factor of $4.6$ with respect to the natural bunch length. The main contribution to the induced voltage in the \gls*{3HC} comes from its impedance at the fundamental harmonic $3\wrf$. The result computed with \gls*{FD} framework taking only the $3\wrf$ mode into account has an agreement of $\Delta\left(\lambda_{\text{A}}, \lambda_{\text{\gls*{FD}}}^{3\wrf}\right)\approx\SI{7e-12}{}$, which was expected given that both models are very similar in terms of the approximations involved. For the result from implementation with \gls*{DFT}, the agreement is $\Delta\left(\lambda_{\text{A}}, \lambda_{\text{\gls*{FD}}}^{\text{DFT}}\right)\approx\SI{1.5e-3}{}$ and for the result from \gls*{SD} method $\Delta\left(\lambda_{\text{A}}, \lambda_{\text{\gls*{SD}}}\right)\approx\SI{1.7e-3}{}$. The equivalence between all equilibrium bunch profiles for uniform fill was confirmed in our results even though it was never assumed \emph{a priori}.

A small systematic difference between the \gls*{SD} and \gls*{FD} frameworks was observed, with value of $\Delta\left(\lambda_{\text{\gls*{SD}}}, \lambda_{\text{\gls*{FD}}}^{\text{DFT}}\right)\approx\SI{3e-4}{}$. In our tests, this difference seems to be insensitive to the number of points considered in the $z$ discretization and its origin is not clear. Nevertheless, we believe that this level of disagreement between methods is too small to have a considerable impact for practical purposes.

The computation time is also an important metric for the comparison of the several approaches discussed here. Considering the hardware described previously, the evaluation of the analytical formula for the induced voltage took $\SI{20}{\milli\second}\slash \text{step}$. For the \gls*{FD} framework, the method of selecting the most relevant harmonics (only one, in this case) was slightly faster ($\SI{70}{\milli\second}\slash\text{step}$) than the implementation with \gls*{DFT} ($\SI{100}{\milli\second}\slash \text{step}$). The \gls*{SD} calculation was also quite fast ($\SI{130}{\milli\second}\slash \text{step}$), since just one resonator was considered. The numerical iterations of Anderson's acceleration algorithm contributes to approximately $\SI{270}{\milli\second}\slash \text{step}$. The analytical implementation and the \gls*{FD} with mode selection converged after 56 iterations. The calculation with \gls*{FD} using \gls*{DFT} converged in 52 iterations and for the \gls*{SD} framework convergence was achieved after 72 iterations. Overall, each simulation took less than \SI{30}{\second} to run. It is worth mentioning that the implementation of \gls*{FD} approach with \gls*{DFT} is independent of the number of impedance sources or on the filling pattern, while the other methods are expected have a strong dependency of the computation time on these factors.

\subsubsection{Hybrid filling\label{subsub:hybrid_filling}}

The nonuniform filling pattern considered here\footnote{This specific hybrid filling pattern is used only as a case study. It does not reflect any plan for operation in SIRIUS storage ring.} consists of a high-charge bunch of~$\SI{2}{\milli\ampere}$ at bucket \num{432}, two gaps of 50 buckets ($\SI{100}{\nano\second}$ gap) around it, and the remaining 763 buckets evenly filled to add up to the total current of~$\SI{350}{\milli\ampere}$. In this example, the equilibrium solution was calculated with the \gls*{SD} framework and three different conditions for the \gls*{FD} approach: including 1, 10 and 100 modes in the summation of Eq.~\eqref{eq:wake_voltage_impedance_2}. The relevant harmonics were selected by means of Eq.~\eqref{eq:modes_selection_criteria} as illustrated in Fig.~\ref{fig:impedance_mode_selection},
\begin{figure}
    \centering
    \includegraphics[width=0.48\textwidth]{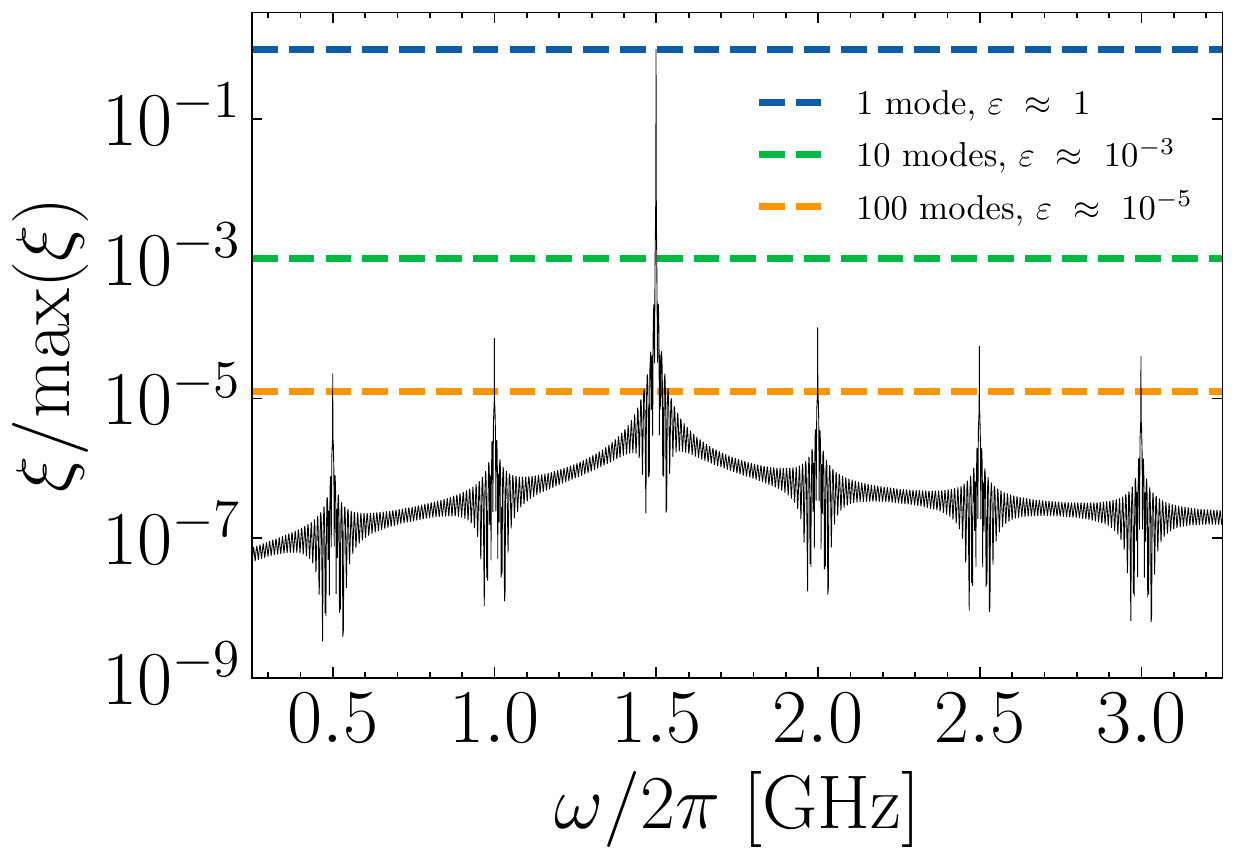}
    \caption{Spectrum $\xi(\omega) = \lvert Z\left(\omega\right) \mathcal{F}\left(\mathbf{I}_b\right) \rvert$ normalized by its maximum value for the hybrid filling pattern and \gls*{3HC} impedance. Horizontal dashed lines represent the thresholds for including 1, 10 and 100 modes.}
    \label{fig:impedance_mode_selection}
\end{figure}
where the normalized spectrum, $\xi(\omega)/\max{\left[\xi(\omega)\right]}$, is shown.  The \gls*{3HC} remained adjusted to the detuning for flat-potential in uniform filling.

As more harmonics are included in the calculation of the beam-induced voltage with the \gls*{FD} framework, it is expected that its results become more similar to the one from the \gls*{SD} framework. This behavior was verified and it is shown in Fig.~\ref{fig:lambda_gap_wake_impedance}.
\begin{figure}
    \includegraphics[width=0.48\textwidth]{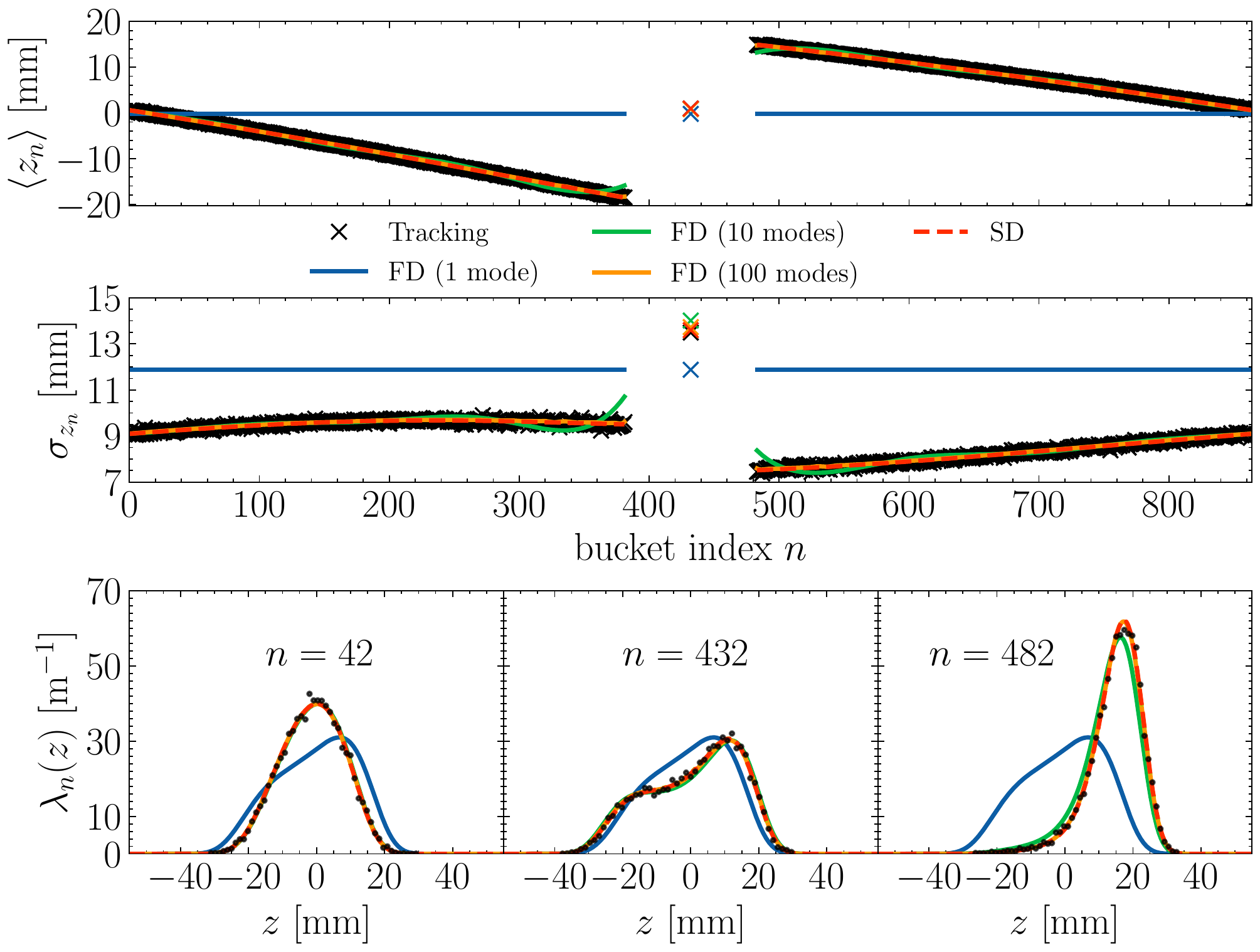}
    \caption{Equilibrium bunch distributions obtained for the hybrid filling pattern with macroparticle tracking, \gls*{SD} and \gls*{FD} approaches for increasing number of modes in the \gls*{FD} framework. The \gls*{3HC} detuning was set to~$\Delta f = \SI{45}{\kilo\hertz}$. Empty buckets are omitted. Bunch centroids~$\langle z_n\rangle$ and bunch lengths~$\sigma_{z_n}$ are shown in the top plot. In the bottom plot, the profiles for three bunches (high-charge bunch in the middle) are compared to tracking results (dots).}
    \label{fig:lambda_gap_wake_impedance}
    \end{figure}
When only 1 mode is included, the beam-induced voltage contains only the contribution from the impedance at $3\wrf$ and the distributions are analogous to the uniform filling case, missing the inhomogeneous beam-loading features. Including \num{10} modes, all of which are revolution harmonics still around $3\wrf$, most of the inhomogeneous pattern is captured. With 100 modes, the criteria defined in Eq.~\eqref{eq:modes_selection_criteria} indicates that other rf harmonics are more relevant to the induced voltage than some revolution harmonics around~$3\wrf$. In this case, the solution from the \gls*{FD} approach shows good agreement with the results obtained from \gls*{SD} method and from tracking as well. The calculation in \gls*{FD} using the \gls*{DFT} approach proved to have the same level of agreement but is not shown in Fig.~\ref{fig:lambda_gap_wake_impedance}. In terms of computing time, the macroparticle tracking simulation for this example took $\SI{3.6}{\hour}$ to run. The calculations with \gls*{SD} framework and the \gls*{FD} approach using \gls*{DFT} reached the equilibrium solution within $\SI{30}{\second}$, the same computing time reported for the uniform filling case. The slowest semi-analytical calculation was with the \gls*{FD} framework considering 100 selected modes, which took $\SI{2}{\minute}$ to reach the equilibrium solution.

\subsection{Effect of llrf feedback\label{subsec:llrf}}

The beam-loading from \glspl*{MC} may have a substantial influence on the longitudinal equilibrium, specially for nonuniform fillings. We used the hybrid filling pattern described in the previous section to illustrate this effect. The equilibrium distributions calculated considering only the passive \gls*{3HC} were compared with the case where the \glspl*{MC} beam-loading is included as well. Different compensation schemes were also tested. In the \gls*{FD} framework\footnote{The implementation with \gls*{DFT} was used to obtain the results from \gls*{FD} framework that are reported in the present and subsequent subsections.} we used the model for the \gls*{llrf} feedback given by Eq.~\eqref{eq:rf_plant_model}. For simplicity, we set the overall gain to $\mathcal{K}=1$. The overall delay considered was~$\tau_\mathrm{d} = \SI{1.9}{\micro\second}$, which is the measured value for the current SIRIUS rf plant. In this scenario, two types of controllers were investigated: one purely integral, with~$k_\mathrm{i}=\SI{0.01}{\ohm^{-1}\second^{-1}}$ and other purely proportional, with~$k_\mathrm{p}=k_\mathrm{p, f}$, where~$k_\mathrm{p, f}~=~\SI{2.96e-6}{\ohm^{-1}}$ is the flat-response gain from Eq.~\eqref{eq:flat_response_gain} for SIRIUS parameters. In the \gls*{SD} framework we applied the phasor compensation scheme. It was verified that the least squares minimization method provided equivalent results, as expected, with the disadvantage of being slower than the phasor method.

The absolute values for the \glspl*{MC} open-loop and closed-loop impedance for each \gls*{llrf} feedback setting are shown in Fig.~\ref{fig:main_cavity_impedance}.
\begin{figure}
    \includegraphics[width=0.45\textwidth]{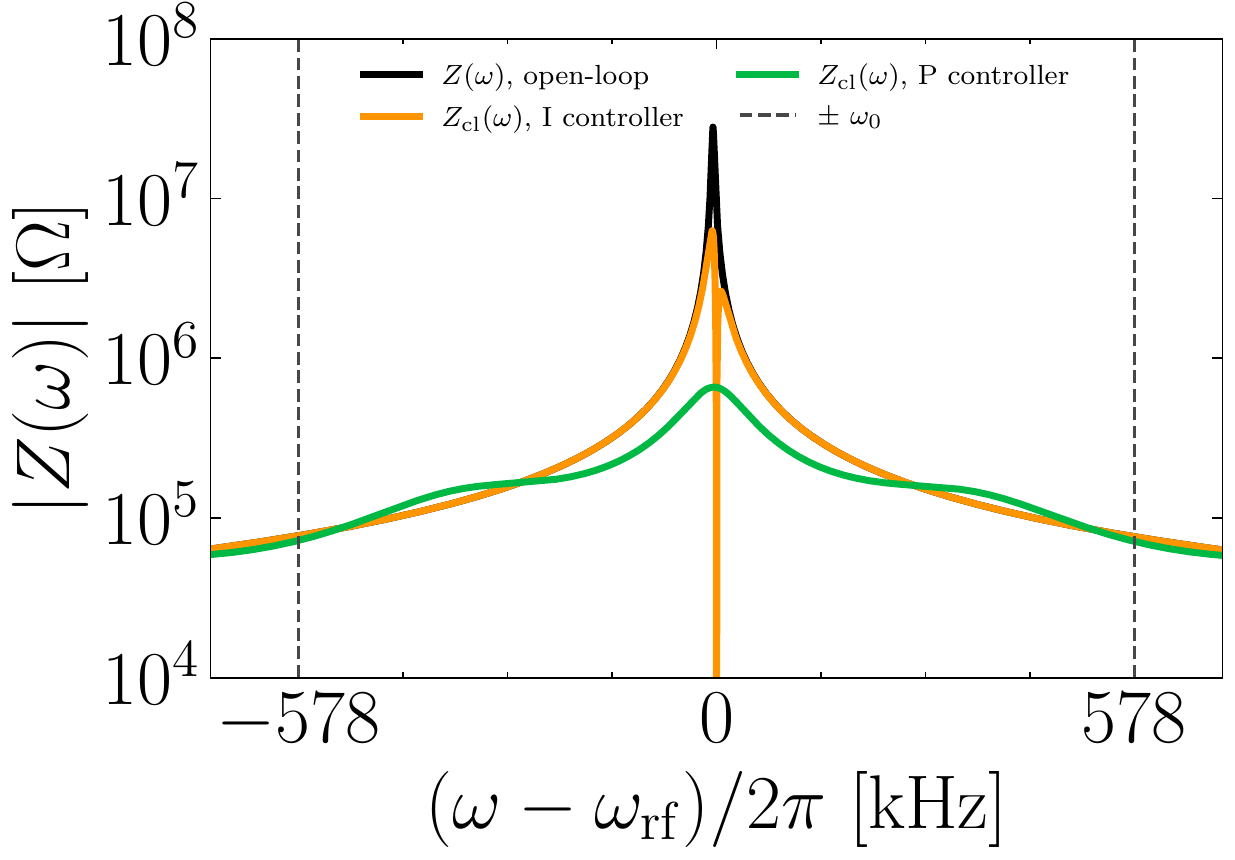}
    \caption{Absolute value of impedance for the \glspl*{MC} in open-loop and closed-loop for two \gls*{llrf} settings, integral (I) and proportional (P) controllers. Revolution harmonics are represented by vertical gray dashed lines.}
    \label{fig:main_cavity_impedance}
\end{figure}
The integral (I) controller heavily suppresses the impedance at the fundamental frequency and acts only on a very narrow bandwidth around it, since a low integrator gain was chosen. The proportional (P) controller does not compensate the beam-loading contribution from $\wrf$ perfectly, but it does reduce the absolute impedance in a considerably broad range of frequencies around $\wrf$. For the cases presented here, neither controller has considerable influence at frequencies of neighboring revolution harmonics. This is commonly the case when only digital \gls*{llrf} are used to control the generator voltage, due to the action of low-pass filters on the measured cavity signal. However, when fast proportional analog feedbacks or more complex topologies are used, the \gls*{llrf} system may impact the beam equilibrium and also the beam stability through its influence on the impedance close to revolution harmonics~\cite{Baudrenghien2017, Karpov2019, Boussard1985}.

Figure~\ref{fig:llrf_effect}
\begin{figure}
    \includegraphics[width=0.48\textwidth]{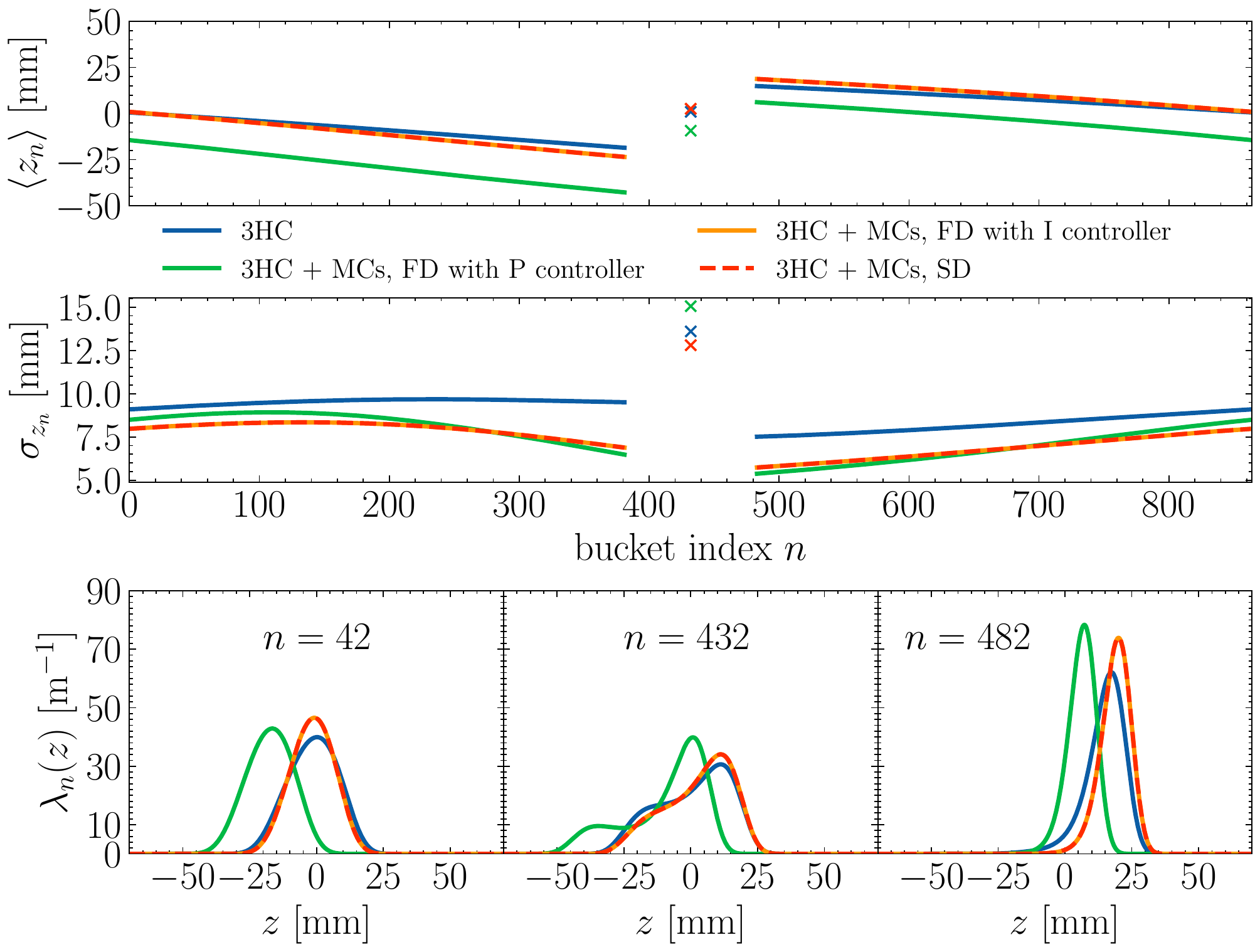}
\caption{Equilibrium bunch distributions for the hybrid filling pattern for different impedance configurations. \gls*{3HC} only (blue) and \gls*{3HC} plus active \glspl*{MC} with beam-loading compensation in three scenarios: with \gls*{llrf} parameters in \gls*{FD} with proportional (green) and integral controller (orange), and phasor compensation in \gls*{SD} (dashed red). Bunch profiles for three buckets are shown in the bottom plot.}
    \label{fig:llrf_effect}
\end{figure}
shows the results for equilibrium bunch centroids and rms bunch lengths for all cases studied. It is clear the equivalence between the solution obtained with \gls*{SD} framework and the one from \gls*{FD} approach using a purely integral controller. We also note that, with the inclusion of \glspl*{MC} beam-loading, the bunch lengthening is systematically reduced for all buckets. With a proportional gain on \gls*{llrf} feedback, the absolute value for the \gls*{MC} closed-loop impedance at rf frequency is about $\SI{1}{\mega\ohm}$. The residual real part of this impedance causes an additional energy loss which induces a negative shift on all bunch centroids and its imaginary part slightly changes the rms bunch length along buckets.
\subsection{Broadband impedance\label{subsec:broadband}}

Figure~\ref{fig:sirius_broadband}
\begin{figure}
    \centering
    \includegraphics[width=0.48\textwidth]{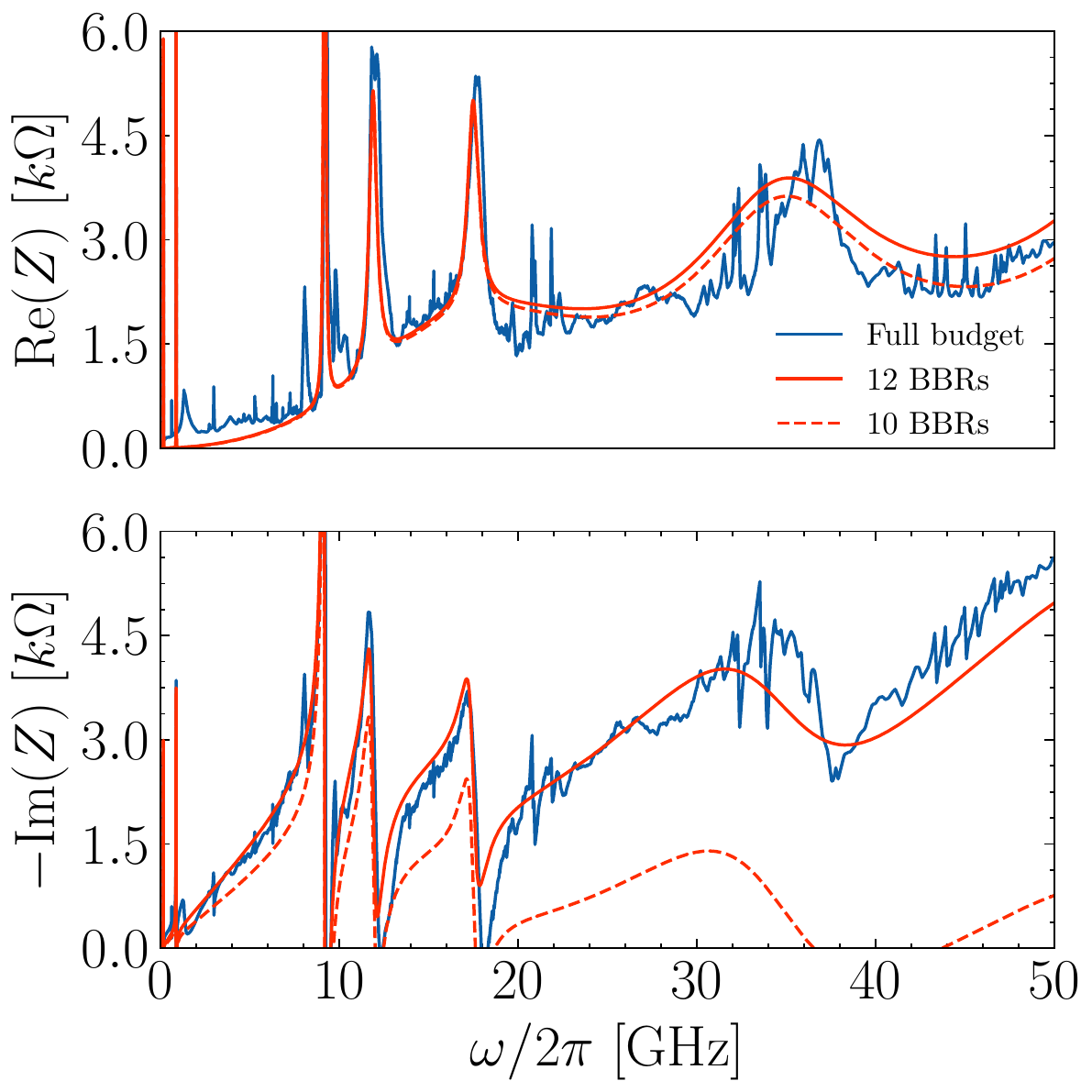}
    \caption{SIRIUS longitudinal impedance for the full budget, 12 BBRs and without the first two BBRs from Table~\eqref{tab:fitted_bbr}.}
    \label{fig:sirius_broadband}
\end{figure}
shows the model of the longitudinal impedance budget of the SIRIUS storage ring and a fitting done with several \glspl*{BBR}, whose parameters are listed in Table~\ref{tab:fitted_bbr}
\begin{table}
    \caption{Fitted parameters to capture the main features of SIRIUS longitudinal impedance budget.}
    \label{tab:fitted_bbr}
\begin{ruledtabular}
    \begin{tabular}{rrrr}
        $f_R$~$\left[\SI{}{\giga\hertz}\right]$ & $R_s$~$\left[\SI{}{\kilo\ohm}\right]$ & $Q$ & $\omega_R/2Q$~$\left[\SI{}{\giga\hertz}\right]$       \\
        \hline
        716.2 & 30.0 & 0.7 & 3216.1 \\
        206.9 & 6.5  & 1.3 & 500.0  \\
        138.4 & 2.0  & 4.0 & 108.7  \\
        79.6  & 2.0  & 1.0 & 250.1  \\
        57.3  & 2.5  & 4.5 & 40.0   \\
        35.0  & 2.5  & 3.0 & 36.6   \\
        17.8  & 1.7  & 1.0 & 55.9   \\
        17.5  & 3.0  & 24  & 2.29   \\
        11.9  & 4.0  & 24  & 1.56   \\
        9.2   & 20.0 & 100 & 0.29   \\
        0.9   & 7.0  & 261 & 0.011  \\
        0.2   & 6.0  & 263 & 0.002
    \end{tabular}
\end{ruledtabular}
\end{table}
~\cite{Sa2018}. The bellows and \gls*{BPMs} are the main contributors for the real part of the impedance and the second most relevant sources to the imaginary part. The narrow peaks at frequencies close to~\SI{9}{\giga\hertz} and~\SI{12}{\giga\hertz} are related to trapped modes in the bellows cavity and the broader peak around~\SI{18}{\giga\hertz} is due to  \gls*{BPMs}. SIRIUS vacuum chamber is mostly composed of a copper cylindrical tube with \SI{12}{\milli\meter} of inner radius and coated with \gls*{NEG}~\cite{Seraphim2015}. The finite resistivity of these chambers is responsible for most of the imaginary part of the impedance budget. This feature is captured by the \gls*{BBR} fitting via the first two resonators from Table~\ref{tab:fitted_bbr}, which have a very high resonant frequency and low quality factor. Figure~\ref{fig:sirius_broadband} highlights the contribution of these two resonators to the overall fitting.

The inclusion of a full broadband impedance model in the \gls*{FD} framework is straightforward. It is sufficient to get the impedance of each contribution at the revolution harmonics, add them and use Eq.~\eqref{eq:wake_voltage_impedance_2} to calculate the total beam-induced voltage. Besides, the computational time in this case does not depend on the number of sources. On the other hand, one possible way of achieving the same result for the \gls*{SD} formulation is to use \gls*{BBR} models to fit the impedance, calculate the induced voltage for each one of them using Eq.~\eqref{eq:wake_voltage_compact} and then sum the contributions. Drawbacks of this procedure include not capturing the exact impedance budget, having a time complexity linear with the number of \glspl*{BBR} and invoking non-physical constructions to represent a physical impedance source. As an example, take the first two \glspl*{BBR} of Table~\ref{tab:fitted_bbr}, which do reproduce the inductive impedance of the resistive-wall wake at low frequencies, but have no physical connection to the original impedance source. Besides, these high frequency resonators are somewhat difficult to simulate due to the numerical problems discussed. Another method to include the effect of broadband impedance in the \gls*{SD} framework is to directly convolve the total wake-function with each bunch distribution~\cite{He2021, Warnock2021b}. This approach, however, would not be correct for wakes that span over a few buckets, such as the ones captured by the last three \glspl*{BBR} of Table~\ref{tab:fitted_bbr}. Even though a combination of the previous methods can be employed, or even other well-known impedance models can be used to fit the budget (such as a purely inductive wake), there is no elegant and simple way of including broadband impedances when using the \gls*{SD} framework.

The effect of the SIRIUS broadband impedance on equilibrium is presented in Fig.~\ref{fig:lambda_with_broadband},
\begin{figure}
    \includegraphics[width=0.45\textwidth]{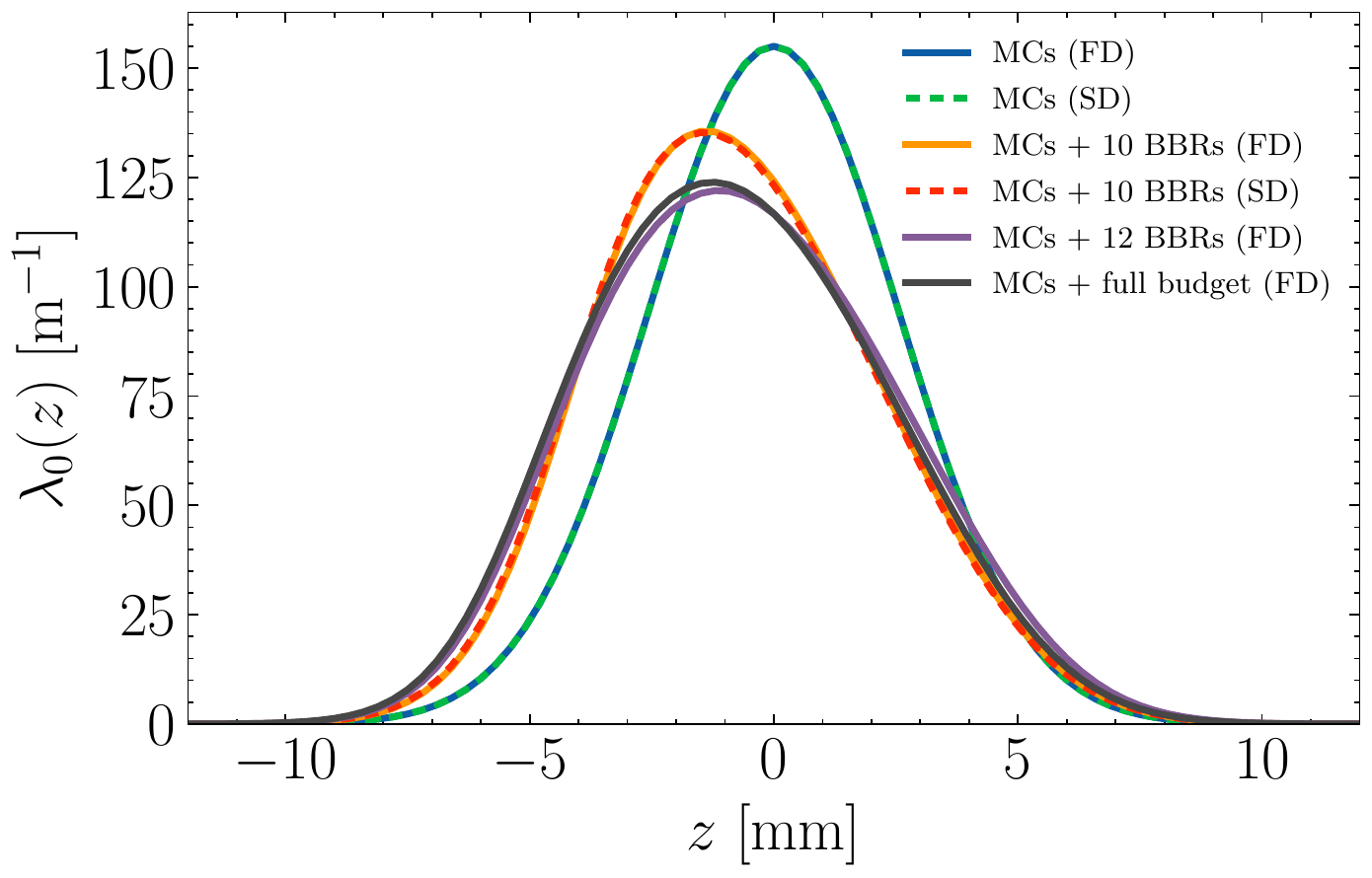}
    \caption{Effect of broadband impedance on longitudinal bunch distribution for uniform filling in the presence of \glspl*{MC}. All results from \gls*{FD} were computed with the \gls*{DFT} implementation.}
    \label{fig:lambda_with_broadband}
\end{figure}
where we simulated the case for nominal current in uniform filling, in the presence of \glspl*{MC} with beam-loading compensation by an integral controller. The \gls*{3HC} will be essential for reaching the nominal current in the real machine due to components heating issues, but we decided to not include it in this simulation to highlight the effect of the broadband impedance on the beam. Considering the bunch length from Table~\ref{tab:sirius_params}, the beam interacts with the impedance up to approximately~$\SI{40}{\giga\hertz}$. However, with a bunch lengthening factor of 4 provided by the \gls*{3HC}, the spectrum would have considerable power only up to~$\SI{10}{\giga\hertz}$. Therefore, it is expected that with a \gls*{3HC}, the broadband impedance impact on the equilibrium and even on time-dependent effects will be reduced.

For the simulations with the \gls*{SD} framework, we did not include the first two \glspl*{BBR} from Table~\ref{tab:fitted_bbr}. Their high resonant frequency would require a much finer discretization of the $z$ domain than the one we used throughout this section. Additionally, the absence of these high frequency resonators helps to emphasize the advantages of the \gls*{FD} framework over the \gls*{SD} approach and their effect on the bunch distribution. We note that the \gls*{FD} simulation with all 12 \glspl*{BBR} is sufficient to reproduce the bunch profile obtained with the full budget, which confirms that the fitting does capture the main features of the impedance. This example indicates that at nominal current the SIRIUS impedance budget would increase the natural bunch length by~$18\%$, from~$\SI{2.57}{\milli\meter}$ to~$\SI{3.04}{\milli\meter}$, cause a shift of~$\SI{-0.74}{\milli\meter}$ in the bunch centroid and make the bunch profile more asymmetric. This rather small bunch lengthening would not sufficiently reduce the heating load at design current, which justifies the need for a \gls*{HHC}.

\subsection{Touschek lifetime improvement with a 3HC\label{subsec:touschek}}

In this last example, the bunch lengthening provided by the superconducting passive \gls*{3HC} that is planned to be installed in SIRIUS storage ring will be studied for some filling patterns. With the \gls*{FD} framework we obtained the longitudinal equilibrium for a beam at nominal current, in the presence of the full broadband impedance budget, a passive \gls*{3HC} and two active \glspl*{MC}. The beam-loading compensation was simulated with a \gls*{llrf} feedback with \gls*{PI} controller. The proportional gain was set to the flat-response value and the integrator gain was adjusted to $k_\text{i}=\SI{0.01}{\ohm^{-1}\second^{-1}}$. Equilibrium distributions were calculated for \num{21} sequentially decreasing \gls*{3HC} detunings from $\SI{50}{\kilo\hertz}$ to $\SI{30}{\kilo\hertz}$, taking the solution from the previous detuning as the initial condition for the next one. All calculations took about $\SI{10}{\min}$ to run and no convergence issues were experienced.

The Touschek loss rate is proportional to the integrated square of the longitudinal bunch distribution. Hence, the relative difference in Touschek lifetimes for two distributions $\lambda_\mathrm{a}(z)$ and $\lambda_\mathrm{b}(z)$ can be calculated as~\cite{Byrd2001}:
\begin{equation}
    \frac{\tau_\mathrm{b}}{\tau_\mathrm{a}} \approx \frac{\int \der z\lambda_\mathrm{a}^2(z)}{\int \der z\lambda_\mathrm{b}^2(z)},
    \label{eq:lifetime_ratio}
\end{equation}
where it is assumed that other parameters that affect Touschek lifetime are the same for the two cases.

Figure \ref{fig:scan_3hc_full_model}
\begin{figure*}
    \includegraphics[width=\textwidth]{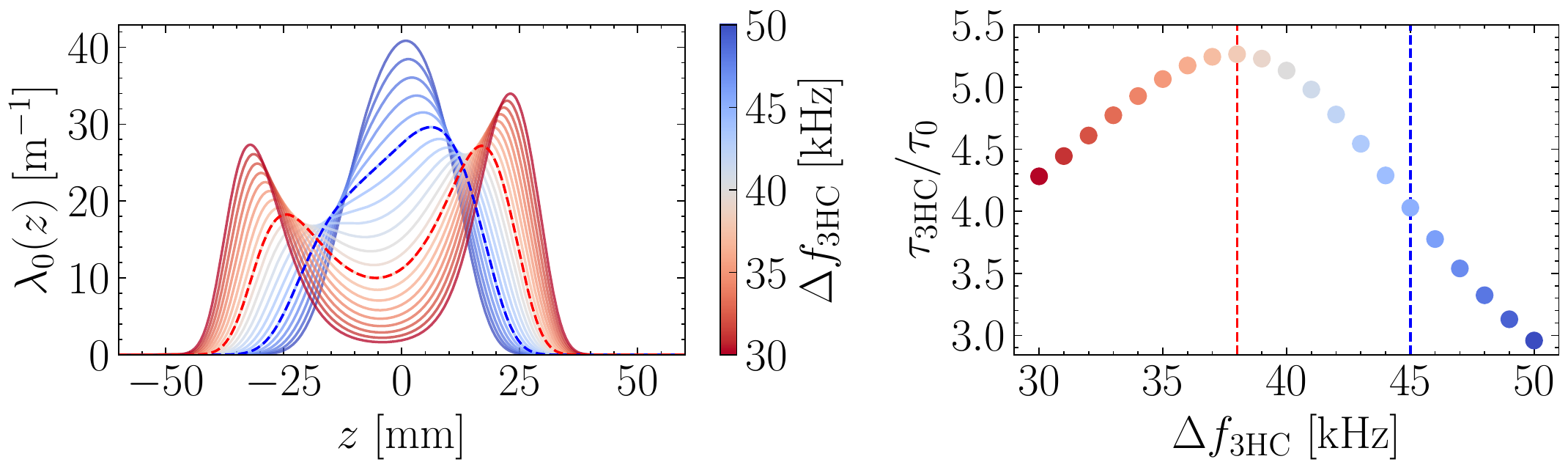}
    \caption{Longitudinal bunch distribution for different \gls*{3HC} detunings (left) and the relative lifetime improvement (right) with respect to the case without the \gls*{3HC}, represented by $\tau_0$. The harmonic voltage for flat-potential condition was obtained with $\Delta f_{\mathrm{3HC}} = \SI{45}{\kilo\hertz}$ and the corresponding distribution is highlighted as a dashed blue curve. The maximum lifetime improvement was obtained with $\Delta f_{\mathrm{3HC}} = \SI{38}{\kilo\hertz}$ and the bunch profile for this condition is emphasized as a dashed red curve.}
    \label{fig:scan_3hc_full_model}
\end{figure*}
shows the bunch distributions on the left and the Touschek lifetime increase with respect to the case without the \gls*{3HC} on the right, for the simulated \gls*{3HC} detunings. It was observed that the \glspl*{MC} beam-loading have a negligible effect on the equilibrium at uniform filling. This was expected due to the heavy suppression of the \glspl*{MC} impedance at~$\wrf$ provided by the \gls*{llrf} feedback. Moreover, the influence of the broadband impedance is also reduced in the presence of the \gls*{3HC}, as discussed in the previous section. With these considerations, the longitudinal bunch distribution is determined mostly by the combination of the generator voltage and the \gls*{3HC} beam-induced voltage. An interesting result is that the maximum lifetime improvement of \num{5.3} happens at~$\Delta f_{\mathrm{3HC}} = \SI{38}{\kilo\hertz}$, while at flat-potential condition, a factor \num{4} is expected. At this optimal condition for lifetime, the bunch is overstretched and the peak harmonic voltage is~$\SI{1.03}{\mega\volt}$, which is $8\%$ higher than the flat-potential voltage.

Figure~\ref{fig:full_model_camshaft}
\begin{figure}
    \includegraphics[width=0.48\textwidth]{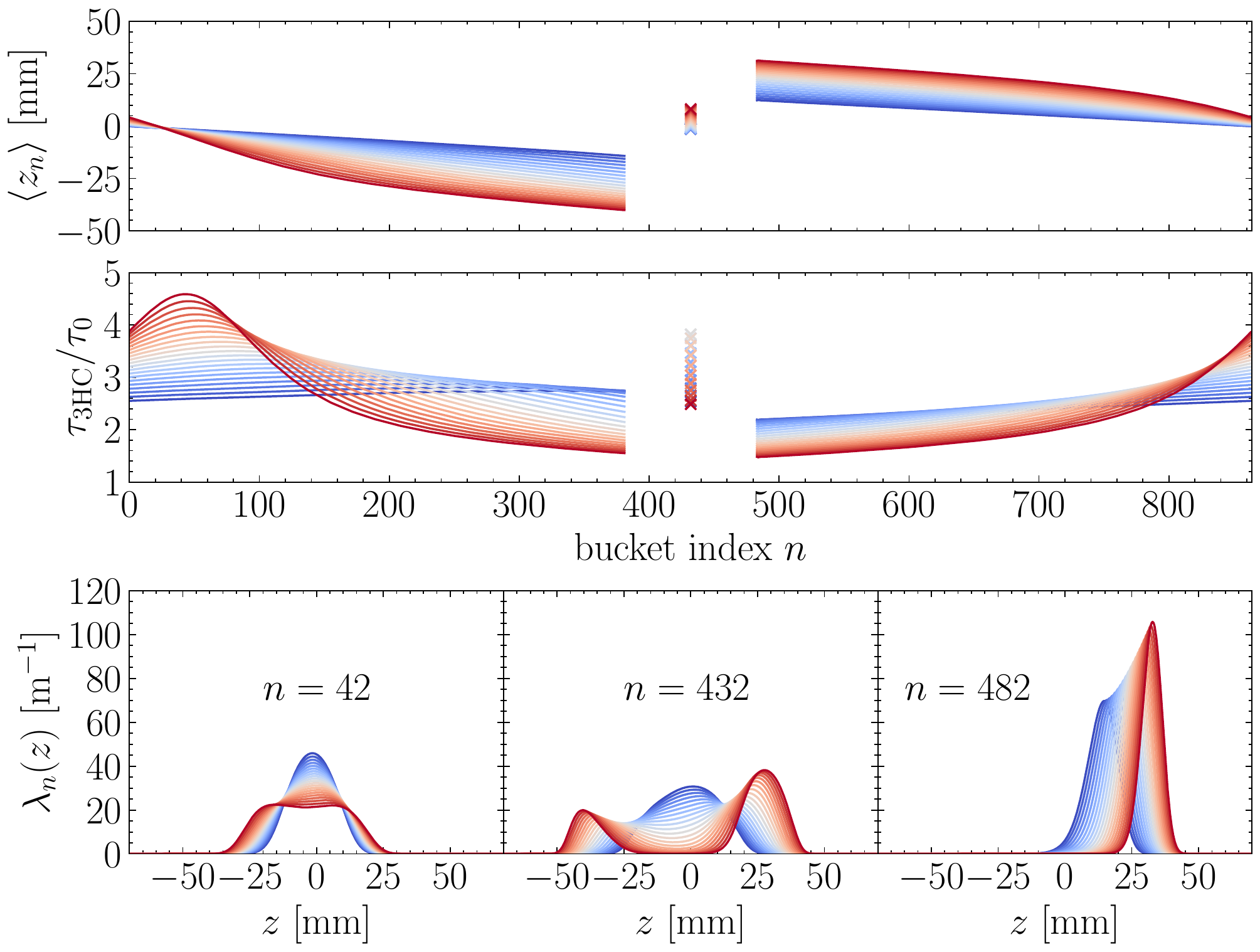}
\caption{Bunch centroids and lifetime improvement factor for the hybrid filling pattern with the full impedance model for SIRIUS. The colors indicate different \gls*{3HC} detunings, following the frequency values from Fig.~\ref{fig:scan_3hc_full_model}. Bunch profiles for three bunches are shown in the bottom plot.}
    \label{fig:full_model_camshaft}
\end{figure}
shows the equilibrium results for the hybrid filling pattern and the same set of \gls*{3HC} detunings from Fig.~\ref{fig:scan_3hc_full_model}. The lifetime improvement factor\footnote{
    We plot the Touschek lifetime improvement factors in Fig.~\ref{fig:full_model_camshaft} because the rms bunch length is not an appropriated metric for overstretched distributions, since the bunch profile is a composition of two shorter bunches.}
was calculated with respect to the case without the \gls*{3HC} and the same filling pattern. Note that for lower detunings the inhomogeneous beam-loading effects are more pronounced. The lifetime ratio is better for bunches in the middle of the train (bucket indices~\numrange{820}{120}). Large bunch centroid shifts and degradation of bunch lengthening for bunches closer to the gaps is observed. The lifetime improvement for the high-charge bunch in the center (index \num{432}) showed a similar qualitative behavior with the reduction of the \gls*{3HC} detuning, as presented in the right plot of Fig.~\ref{fig:scan_3hc_full_model}. From these results it is clear that simply reducing the \gls*{3HC} detuning is an ineffective approach to improve the overall lifetime for hybrid filling patterns and other strategies should be employed. A better solution can be the introduction of guard bunches to compensate the inhomogeneous beam-loading caused by gaps~\cite{Milas2010, Olsson2018, Warnock2020, He2021}.

\section{Conclusion\label{sec:conclusion}}
In this paper, we derived two approaches to compute the equilibrium beam-induced voltage in the presence of arbitrary filling patterns and impedance sources. The calculation in \gls*{SD} framework is limited to resonator wake-functions. The theory found in the literature~\cite{Olsson2018, Warnock2020, Warnock2021a, Warnock2021b, He2021}, was revisited, extended to consider the most general resonator model and formulated in a compact equation, convenient for numerical implementation in a uniformly discretized grid. A different approach, based on the \gls*{FD} analysis, allowed the generalization upon arbitrary impedance sources and offered a straightforward process for computing the beam-induced voltage in terms of \glspl*{DFT}. The low computational cost of the \gls*{FD} framework is noteworthy, as it has the benefits from \gls*{FFT} algorithms and its time complexity is constant besides the number of impedance elements. We benchmarked the results using the parameters of SIRIUS storage ring, a fourth-generation synchrotron. For uniform filling and narrowband resonators, it was analytically and numerically demonstrated that the two proposed frameworks reduce to a well-known formula for the beam-induced voltage. For nonuniform filling, the methods were benchmarked against macroparticle tracking and the results exhibited excellent agreement.

The beam-loading compensation of active rf cavities was addressed with the concept of closed-loop impedance. This approach can only be applied in the \gls*{FD} framework and is conceptually different from other methods based on phasor compensation or least square minimization. In the latter the parameters of the external voltage are adjusted to compensate the beam-loading, while the former changes the impedance model of the cavity so that the beam-induced voltage is intrinsically compensated. We observed that the stationary beam-loading compensation methods as described in Ref.~\cite{Warnock2021a} are equivalent to a closed-loop impedance of an integral controller with low gain. The proposed approach allows more realistic simulations of active rf cavities and is flexible to model several \gls*{llrf} system topologies.

Another advantage of the \gls*{FD} over the \gls*{SD} framework was illustrated with the simulation of the SIRIUS broadband impedance budget. This was easily accomplished in \gls*{FD} by taking the full impedance budget as a direct input for the calculations. In contrast, in \gls*{SD} the inclusion of broadband impedance requires additional steps, such as fitting \glspl*{BBR} or convolving short-range wake-function with longitudinal distributions~\cite{Warnock2021a, He2021, Warnock2021b}. These approaches, however, may introduce several numerical issues that must be handled and typically require a case-by-case analysis to define how each impedance contribution should be simulated.

We also studied the effect of different detunings of a passive superconducting \gls*{3HC} on Touschek lifetime, taking into account the complete impedance budget for SIRIUS storage ring. For uniform filling, the maximum Touschek lifetime improvement was obtained with an overstretched bunch profile, increasing it by a factor \num{5.2}, while the flat-potential condition is expected to the increase lifetime by a factor of \num{4}. The more involved case of nonuniform filling pattern was briefly discussed only to illustrate the flexibility of the tool. Further investigations and more accurate metrics should be considered to compare performances in this case.

It is important to mention that the existence of equilibrium in simulation does not imply stability in the real machine. The map for reaching the steady-state in simulations is based on robust fixed-point algorithms, while the real dynamics depends on the intricate balance between damping and coherent excitation. As an example, to provide bunch lengthening, \glspl*{HHC} must operate at ac Robinson unstable detunings~\cite{Ng2005}. Fortunately, \glspl*{MC} can often be adjusted to a Robinson stable detuning and provide enough damping. However, for small \gls*{HHC} detunings as presented in this paper, this balance should be carefully checked. Other time-dependent effects introduced by \glspl*{HHC} that can limit the achievable bunch lengthening may include, in particular, the recently predicted~\cite{Venturini2018} and observed~\cite{Cullinan2022talk} mode-1 instability and, more generally, some instability induced by the reduction of the average incoherent synchrotron frequency as the longitudinal potential is flattened. A complete study covering time-dependent effects was beyond the scope of this work. Nevertheless, the developed framework can be useful in such studies for computing the unperturbed bunch distributions, which are essential inputs for single-bunch and multi-bunch instability thresholds calculation~\cite{Venturini2018, Lindberg2018, Cullinan2020, Cullinan2022}.

In summary, the proposed \gls*{FD} methods proved to be more general, numerically stable and faster than the \gls*{SD} framework. This makes it a helpful tool during the design phase of a storage ring, when different specifications are being explored and the impact of machine components impedance on beam parameters should be quantified.

\begin{acknowledgments}
The authors would like to thank the \gls*{LNLS} rf group for fruitful discussions on \gls*{llrf} feedback loops and the members of \gls*{LNLS} accelerator physics group for careful reading of the manuscript and valuable suggestions.
\end{acknowledgments}

\appendix*
\section{Limit case of uniform filling and passive narrowband resonator\label{app:uniform_fill}}
In this Appendix we will apply the equations derived in Sec.~\ref{sec:theory} to check its limit for a specific scenario: uniform filling pattern and high-$Q$ resonator.

Consider the case of~$h$ bunches in a ring evenly filled with the same current per bunch~$I_{\ell} = I_\mathrm{t}/h$. In the equilibrium state, the longitudinal distributions and beam-induced voltage will be equivalent for all bunches. Without loss of generality, we will take the rf bucket 0 as reference for the derivation.

\subsection{Frequency-domain}
Applying the uniform filling considerations to Eq.~\eqref{eq:wake_voltage_impedance_2} reads
\begin{eqnarray}
    V_0(z) &=& -2(I_\mathrm{t}/h)\Real\Biggl[\sum_{p=0}^{+\infty}Z^\ast( p\omega_0)e^{ip\omega_0z/c}    \nonumber \\
           & & \times~\hat{\lambda}_{0}^\ast(p\omega_0)\sum_{\ell=0}^{h-1}e^{-2\pi i p\ell/h}\Biggr].
    \label{eq:full_voltage_impedance}
\end{eqnarray}

The geometric series sum over~$\ell$ yields
\begin{equation}
    \sum_{\ell=0}^{h-1}e^{- 2\pi i p \ell/h} = h\delta_{p, qh} \,\, \text{for}~q \in \mathbb{N}
    \label{eq:sum_over_buckets}
\end{equation}
where~$\delta_{p, qh}$ is the Kronecker delta. Thus, the beam samples the impedance only at rf harmonics, which is expected from the symmetry of uniform filling.

Assuming a high-$Q$ narrowband resonator impedance sharply peaked close to the $m$th rf harmonic, the major contribution to the beam-induced voltage is related to the term~$q=m$. Applying Eq.~\eqref{eq:sum_over_buckets} into Eq.~\eqref{eq:full_voltage_impedance} and retaining only the contribution from $\omega_m:=m\wrf$:
\begin{equation}
    V_0(z) = -2I_\mathrm{t}\Real\left[Z^\ast(\omega_m)\hat{\lambda}_{0}^\ast(\omega_m)e^{i\omega_m z/c}\right].
    \label{eq:voltage_partial}
\end{equation}

A convenient parametrization for the Fourier transform of longitudinal bunch distribution is
\begin{equation}
    \hat{\lambda}_0(\omega) = F_0(\omega)e^{i\Phi_0(\omega)}
    \label{eq:form_factor}
\end{equation}
where, to respect the property $\hat{\lambda}_0(-\omega) = \hat{\lambda}_0^{*}(\omega)$, $F_0(\omega)$ must be a real-valued even function and $\Phi_0(\omega)$ a real-valued odd function. With this parametrization, Eq.~\eqref{eq:voltage_partial} can be arranged as
\begin{equation}
    V_0(z) = -2I_\mathrm{t}F_0(\omega_m)\Real\left[Z^\ast(\omega_m)e^{i\left[\omega_m z/c - \Phi_0(\omega_m)\right]}\right].
\end{equation}

The model for resonator impedance is given by the RLC circuit impedance from Eq.~\eqref{eq:rlc_impedance}, rewritten as:
\begin{align}
    Z(\omega) = \dfrac{R_s}{1+ 2 i Q \delta_\omega},
\end{align}
with the resonator relative detuning defined as
\[
\delta_\omega \coloneqq \frac{1}{2}\left(\frac{\omega_R}{\omega}  - \frac{\omega}{\omega_R}\right).
\]

For frequencies close to resonance $\omega \approx \omega_R$, the approximated formula $\delta_\omega = \Delta \omega \slash \omega_R$ with $\Delta \omega = \omega_R - \omega$ is commonly used. The present analysis refers to $\omega_m \approx \omega_R$.

With the RLC impedance model, we are able to cast the wake voltage in the form:
\begin{equation}
    V_0(z) = -2I_\mathrm{t} R_s\dfrac{F_0(\omega_m)}{1+4Q^2\delta^2_\omega} \left[\cos\theta_0(z) - 2Q\delta_\omega \sin\theta_0(z)\right],
    \label{eq:partial_voltage_unifill}
\end{equation}
where $\theta_0(z) = \omega_m z/c - \Phi_0(\omega_m)$.

Defining
\begin{equation}
    \tan \psi \coloneqq 2 Q \delta_\omega,
    \label{eq:detuning_angle}
\end{equation}
then Eq.~\eqref{eq:partial_voltage_unifill} can be further simplified to
\begin{eqnarray}
    V_0(z) &=& -2 I_\mathrm{t} F_0(\omega_m) R_s \cos \psi \nonumber \\
        & & \times \cos\left[\omega_m z\slash c + \psi - \Phi_0(\omega_m)\right]
        \label{eq:analytic_voltage}
\end{eqnarray}
which is a well-known formula for the equilibrium beam-induced voltage in uniform filling with a passive narrowband resonator, including the so-called complex bunch form-factor~\cite{Tavares2014}.

\subsection{Space-domain}
For high-$Q$ resonators, we shall consider that the $I_0S_0(z)$ contribution in Eqs.~\eqref{eq:k_function} and~\eqref{eq:k_function2} is negligible as compared to the summation part, since $S_0(z)$ only accounts for the self-induced voltage of a bunch on itself on the present turn. In this case and considering uniform filling, Eq.~\eqref{eq:k_function2} can be simplified to
\begin{equation}
    K_0(z) = (I_\mathrm{t}/h)e^{-\kappa z}\frac{S_0(\lambda_\rf/2)}{1-\nu} \sum_{\ell=1}^{h}\nu^{\ell/h}
    \label{eq:kfunc_uniform_filling}
\end{equation}

We can approximate $\alpha/c \ll 1$ for long-range wakefields and take $e^{-\alpha z/c} \approx 1$. Moreover, since $\alpha/\omega_R=1/2Q \ll 1$, it follows that $\wrb \approx \omega_R$. Therefore, $e^{\pm\kappa z}~\approx~e^{\mp i\omega_R z/c}$. In this scope, the Laplace transform can be replaced by the Fourier transform:
\[
    S_0(\lambda_\rf/2) = \int_{-\lambda_\rf/2}^{\lambda_\rf/2}{\der z'\lambda_0(z')}e^{\kappa z'} \approx \hat{\lambda}_0^{\ast}(\omega_R).
\]

The sum over bunches~$\ell$ is the sum of $h-1$ terms of a geometric series with common ratio $\nu^{1/h}$, hence:
\begin{equation*}
\sum_{\ell=1}^{h}\nu^{\ell/h} = \nu^{1/h}\frac{1-\nu}{1-\nu^{1/h}}
\end{equation*}

Recall that $\nu^{1/h} = e^{-\kappa C_0/h} = e^{-\kappa \lambda_\rf}$. Applying those partial results into Eq.~\eqref{eq:kfunc_uniform_filling} yields:
\begin{equation}
    K_0(z) = (I_\mathrm{t}/h)e^{ i\omega_R z/c} \frac{\hat{\lambda}_0^{\ast}(\omega_R)}{e^{\kappa \lambda_\rf} - 1}.
    \label{eq:kfunction_zero_partial}
\end{equation}

Following the method applied in Ref.~\cite{Warnock2020}, let $(\rho, \psi)$ be polar variables in the complex plane such that
\begin{equation}
    \dfrac{1}{e^{\kappa \lambda_\rf} - 1} = \rho e^{i\psi}.
\end{equation}

With $\lambda_\rf = 2\pi c/\wrf $, then $e^{\kappa \lambda_\rf}~=~e^{2\pi \alpha / \wrf }e^{-2\pi i \omega_R / \wrf }$. Considering that the resonant frequency is close to the $m$th rf harmonic, the detuning is $\Delta \omega~=~\omega_R - m\wrf $. From this, $e^{-2\pi i \omega_R / \wrf }~=~e^{-2\pi i \Delta \omega / \wrf }$ follows. Besides, assuming a small detuning such that $\Delta \omega / \wrf  \ll 1$, the exponential can be approximated in first order to:
\begin{eqnarray*}
    e^{\kappa \lambda_\rf} - 1 &\approx& \left(1+2\pi \alpha / \wrf \right)\left(1-2\pi i \Delta \omega / \wrf \right) - 1\\
    &\approx& \frac{1}{\frf }\left(\alpha - i\Delta \omega\right),
\end{eqnarray*}
where the second order term proportional to $\alpha \Delta \omega/\wrf^2$ was neglected. Therefore:
\begin{equation}
    \rho = \left|\dfrac{1}{ e^{\kappa \lambda_\rf} - 1}\right| \approx \dfrac{\frf}{\sqrt{\alpha^2 + \Delta\omega^2}},
    \label{eq:absolute_value}
\end{equation}
and the phase can be calculated with
\begin{eqnarray}
    \psi = \arg\left(\dfrac{1}{e^{\kappa \lambda_\rf}  - 1}\right) &=& -\arg\left(e^{\kappa \lambda_\rf}  - 1\right) \nonumber \\
    &=& \arctan\left(\Delta \omega/\alpha\right).
    \label{eq:phase}
\end{eqnarray}

Note that $\tan{\psi}~=~2 Q \Delta\omega /\omega_R$, which is the same relation between the detuning phase $\psi$ and resonator parameters defined in Eq.~\eqref{eq:detuning_angle}.

With those approximations and using Eq.~\eqref{eq:form_factor} for the bunch spectrum, Eq.~\eqref{eq:kfunction_zero_partial} simplifies to
\begin{equation}
    K_0(z) = \frac{(I_\mathrm{t}/h) \frf}{\sqrt{\alpha^2 + \Delta\omega^2}} F_0(\omega_R) e^{i\left[\omega_R z/c + \psi - \Phi_0(\omega_R)\right]}
\end{equation}

Applying this result to~Eq.~\eqref{eq:wake_voltage_compact} reads, after some manipulations,
\begin{eqnarray}
    V_0(z) &=& -2 I_\mathrm{t} F_0(\omega_R)R_s \cos\psi  \nonumber \\
           & & \times~\left[\cos\gamma_0(z) - \frac{\alpha}{\omega_R}\sin\gamma_0(z)\right],
    \label{eq:wake_voltage_partial}
\end{eqnarray}
where $\gamma_0(z) = \omega_R z / c + \psi - \Phi_0(\omega_R)$.

The sine term in Eq.~\eqref{eq:wake_voltage_partial} can be neglected since $\alpha/\omega_R \ll 1$. Additionally, approximating the resonant frequency to its closest rf harmonic $\omega_R \approx \omega_m$, then Eq.~\eqref{eq:wake_voltage_partial} is equivalent to the formula in Eq.~\eqref{eq:analytic_voltage}.

\bibliography{references.bib}
\end{document}